\documentclass[useAMS,usenatbib,usegraphicx]{mn2e}

\usepackage{amssymb}

\newcommand{\Id}{\mathbfss{I}}
\newcommand{\disp}{\mathbb{D}}
\newcommand{\FAP}{{\rm FAP}}
\newcommand{\Res}{{\rm Res}}

\title[Using multi-harmonic periodograms]{Detecting non-sinusoidal periodicities in observational data using multi-harmonic periodograms}
\author[R.V.~Baluev]{Roman V. Baluev\thanks{E-mail: roman@astro.spbu.ru}\\
Sobolev Astronomical Institute, St Petersburg State University, Universitetskij prospekt
28, Petrodvorets, St Petersburg 198504, Russia}

\begin{document}

\date{Accepted 2009 Frbruary 12.
      Received 2009 February 9;
      in original form 2008 November 5}

\pagerange{\pageref{firstpage}--\pageref{lastpage}} \pubyear{2009}

\maketitle

\label{firstpage}

\begin{abstract}
We address the problem of assessing the statistical significance of candidate
periodicities found using the so-called `multi-harmonic' periodogram, which is being used
for detection of non-sinusoidal signals, and is based on the least-squares fitting of
truncated Fourier series. The recent investigation \citep{Baluev08a} made for the
Lomb-Scargle periodogram is extended to the more general multi-harmonic periodogram. As a
result, closed and efficient analytic approximations to the false alarm probability,
associated with multi-harmonic periodogram peaks, are obtained. The resulting analytic
approximations are tested under various conditions using Monte Carlo simulations. The
simulations showed a nice precision and robustness of these approximations.
\end{abstract}


\section{Introduction}
The \citet{Lomb76}-\citet{Scargle82} (hereafter LS) periodogram is a well-known powerful
tool, which is widely used to search for periodicities in observational data. The main
idea used in the LS periodogram is to perform a least-squares fit of the data with a
sinusoidal model of the signal and then to check how much the resulting weighted r.m.s.
have decreased for a given signal frequency. The maximum value of the LS periodogram
(i.e., the maximum decrement in the least-squares goodness-of-fit measure) corresponds to
the most likely frequency of the periodic signal. This natural idea is quite easy to
implement in numerical calculations.

However, random errors in the input data inspire noise peaks on the periodogram, so that
we can never be completely sure that the peak that we actually observed was produced by a
real periodicity. The common way to assess the statistical significance of the observed
peak is based on the associated `false alarm probability' (hereafter $\FAP$). The $\FAP$
is the probability that the observed or larger periodogram peak could be produced by
random measurement errors. The smaller is $\FAP$, the larger is the statistical
significance. Given some tolerance value $\FAP_*$ (say, $1\%$), we could claim that the
detected candidate periodicity is statisticaly significant (if $\FAP<\FAP_*$) or is not
(if $\FAP>\FAP_*$).

From the statistical viewpoint, the $\FAP$ is tightly connected with the probability
distribution of periodogram maxima, which are calculated within some \emph{a priori} fixed
frequency segment.\footnote{Speaking more precisely, the periodogram maxima are always
calculated over some \emph{discrete} set of values. However, in practice the periodograms
are usually plotted on a dense frequency grid, which is practically equivalent to a
continuous segment. It is the case that we consider in the paper.} However, even
approximate calculation of this distribution is a non-trivial task. It represented a
trouble for astronomers for about three decades. It is worthwile to mention here, for
instance, the papers by
\citet{HorneBal86,Koen90,SchwCzerny98a,SchwCzerny98b,Cumming99,Cumming04,Frescura08}.
Recently, a significant progress in this field was attained in the paper
\citep{Baluev08a}, where closed and simultaneously rather efficient approximations of the
$\FAP$ for the LS periodogram are given, basing on results in the theory of extreme values
of stochastic processes.

However, periodic signals being dealt with in astronomy often are significantly
non-sinusoidal. Then the use of the LS periodogram is not optimal, since the corresponding
periodic variation would be fitted inadequately. For instance, it is the case for
lightcurves of variable stars of several types and for radial velocity curves of stars
orbited by a planet on an eccentric orbit. Several ways to deal with this issue were
proposed \citep[for further references see e.g.][]{SchwCzerny98a,SchwCzerny98b}. In this
paper, we focus attention on the so-called multi-harmonic periodogram
\citep{SchwCzerny96}, which is based on the least-squares fitting of truncated Fourier
series. Note that in the paper \citep{Baluev08a} a general class of periodograms based on
the least-squares data fitting was considered as well, but from theoretical positions
only. Here our aim is to apply these general results to the multi-harmonic periodograms.

The plan of the paper is as follows. In Section~\ref{sec_def}, we formulate the problem
rigorously and introduce the necessary mathematical definitions. In Section~\ref{sec_FAP},
basing on the work \citep{Baluev08a}, we derive closed approximations of the $\FAP$,
associated with multi-harmonic periodogram peaks. In Section~\ref{sec_Simul}, we use
numerical Monte Carlo simulations to quantify the accuracy of these analytic
approximations.

\section{General definitions}
\label{sec_def}
Let us write down the temporal model of the putative periodic signal using a trigonometric
polynomial of some \emph{a priori} stated degree $n$:
\begin{equation}
\mu(t,\btheta,f) = \sum_{k=1}^{n} \left( a_k \cos 2\pi kf t + b_k \sin 2\pi kf t \right),
\label{multiharm-model}
\end{equation}
where $f$ is the signal frequency and the vector $\btheta$ incorporates $d=2n$ Fourier
coefficients $a_k, b_k$. Further we adopt exactly the same notations as those used in
\citep{Baluev08a}. Clearly, the model $\mu$ is linear: $\mu(t,\btheta,f) = \btheta \cdot
\bvarphi(t,f)$, where the vector $\bvarphi(t,f)$ incorporate the first $n$ harmonics of
the Fourier basis. In addition to the signal model $\mu$, we define the base temporal
model $\mu_{\mathcal H}(t,\btheta_{\mathcal H}) = \btheta_{\mathcal H} \cdot
\bvarphi_{\mathcal H}(t)$, which is assumed to be linear with respect to $d_{\mathcal H}$
unknown parameters $\btheta_{\mathcal H}$. This base model may represent, for instance, a
constant or a long-term polynomial (e.g., linear or quadratic) temporal trend. Therefore,
the alternative (full) model is given by $\mu_{\mathcal K}(t,\btheta_{\mathcal K},f) =
\mu_{\mathcal H}(t,\btheta_{\mathcal H}) + \mu(t,\btheta,f)$, where $\btheta_{\mathcal K}$
incorporates all parameters in $\btheta_{\mathcal H}$ and $\btheta$. From the viewpoint of
the statistical tests theory, we need to test the base hypothesis $\mathcal H: \btheta=0$
against the alternative one $\mathcal K:\btheta\neq 0$.

The input dataset consists of $N$ measurements $x_i$ taken at timings $t_i$ and having
uncertainties $\sigma_i$. We assume that the random errors of the measurements are
statistically independent and normally distributed. Below we will deal with the
least-squares periodograms defined in \citep{Baluev08a}. These periodograms are based on
the linear least-squares fitting procedure. The basic one, $z(f)$, represents the
half-difference
\begin{equation}
 z(f) = \left[ \chi_{\mathcal H}^2 - \chi_{\mathcal K}^2(f) \right]/2,
\label{PeriodogramDef}
\end{equation}
where $\chi_{\mathcal H}^2$ and $\chi_{\mathcal K}^2$ represent the minimum values of the
$\chi^2$ goodness-of-fit statistic, calculated under the two corresponding hypotheses,
$\mathcal H$ and $\mathcal K$. Note that under the base hypothesis $\mathcal H$ the random
quantities $\chi_{\mathcal H}^2$ and $\chi_{\mathcal K}^2$ follow the
$\chi^2$-distributions with $N_{\mathcal H} = N-d_{\mathcal H}$ and $N_{\mathcal K} =
N-d_{\mathcal K}$ degrees of freedom and thus indeed represent $\chi^2$-variates. The
periodogram $z(f)$ can be only calculated if the variances $\sigma_i$ of the observational
errors are known exactly. Usually we do not know these variances exactly, and can fix only
the statistical weights, $w_i\propto 1/\sigma_i^2$, so that $\sigma_i^2 = \kappa/w_i$ with
the common factor $\kappa$ being unconstrained \emph{a priori}. Therefore, we will also
consider three modified least-squares periodograms:
\begin{eqnarray}
 z_1(f) = N_{\mathcal H} \frac{\chi_{\mathcal H}^2-\chi_{\mathcal K}^2(f)}{2\chi_{\mathcal H}^2}, \quad
 z_2(f) = N_{\mathcal K} \frac{\chi_{\mathcal H}^2-\chi_{\mathcal K}^2(f)}{2\chi_{\mathcal K}^2(f)}, \nonumber\\
 z_3(f) = \frac{N_{\mathcal K}}{2} \ln \frac{\chi_{\mathcal H}^2}{\chi_{\mathcal K}^2(f)}.
\label{PeriodogramModDef}
\end{eqnarray}
These periodograms do not depend on $\kappa$ and can be calculated even if $\kappa$ is
unknown. The periodograms $z_1(f)$ and $z_2(f)$ represent normalizations of the basic
periodogram $z(f)$ by the sample variances of the residuals, calculated under one of the
two hypotheses, $\mathcal H$ or $\mathcal K$. The periodogram $z_3(f)$ is proportional to
the logarithm of the likelihood ratio statistic. More discussion of these definitions can
be found in \citep{Baluev08a}. A discussion of several issues associated with the
least-squares interpretation of the periodograms introduced above can be also found in
\citep{SchwCzerny98a,SchwCzerny98b,ZechKur09}. The modified periodograms $z_{1,2,3}$ are
unique-value monotonic functions of each other and thus are entirely equivalent for the
practical use.

The definitions~(\ref{PeriodogramModDef}) imply that the statistical weights $w_i$ should
be known with sufficient precision, and only the proportionality factor is unknown. This
framework is a usually adopted for the period analysis of astronomical data
\citep[e.g.][]{GillilandBaliunas87,Irwin89,ZechKur09} and we adopt it here. Nevertheless,
sometimes this model may not work well. For instance, the paper \citep{Baluev08b}
discusses the case in which the weights of observations are not known \emph{a priori} with
sufficient precision. In this case, the traditional multi-harmonic periodograms being
discussed here may not work well.

We do not discuss in detail the numerical algorithms for calculation of the periodograms
introduced. The form of the above definitions is more suitable for quantifying the
statistical distributions of the corresponding periodograms. Fast numerical algorithms of
practical evaluation of the multi-harmonic periodograms are given in
\citep{SchwCzerny96,Palmer09}.

\section{False alarm probability}
\label{sec_FAP}
Let us pick any of the periodograms introduced above, and denote it as $Z(f)$. If the
frequency of the putative signal was known, the false alarm probability $\FAP_{\rm
single}(Z)$, associated with the given value $Z(f)$, could be calulated as $\FAP_{\rm
single}(Z)=1-P_{\rm single}(Z)$, where $P_{\rm single}(Z)$ is the cumulative distribution
of the corresponding periodogram value, calculated under the base hypothesis $\mathcal H$.
It is well-known that within simple constant scale factors these distributions are
$\chi^2(d)$, $F(d,N_{\mathcal K})$, and $B(d,N_{\mathcal K})$ for the periodograms $z$,
$z_2$, and $z_1$, respectively \citep[see,
e.g.,][]{SchwCzerny98a,SchwCzerny98b,Baluev08a}. Here the quantities in brackets mark the
necessary numbers of degrees of freedom.

When the signal frequency is unknown \emph{a priori}, we need to search for a maximum of
$Z(f)$ within some wide frequency band $[f_{\rm min},f_{\rm max}]$. From now on we will
assume, for the sake of definiteness, that $f_{\rm min}=0$. In practice it is a frequent
case and also this assumption allows us to simplify the formal expressions. All results
presented below can be easily extended to the case of arbitrary $f_{\rm min}>0$. For
example, we will need to replace certain lower integration limits appropriately and to
change the expressions for the frequency bandwidth from $f_{\rm max}$ to $f_{\rm max} -
f_{\rm min}$. According to \citet{Baluev08a}, to estimate the $\FAP$ associated with the
observed maximum, we use the \citet{Davies77,Davies87,Davies02} bound
\begin{equation}
\FAP_{\rm max}(Z,f_{\rm max}) \leq \FAP_{\rm single}(Z) + \tau(Z,f_{\rm max}),
\label{Dav}
\end{equation}
Exact expressions for function $\tau$ are given in \citep{Baluev08a} for the general
least-squares periodogram $z$ and for its modifications $z_{1,2,3}$ (see eqs.~(7) and~(8)
in that paper). In fact, the right hand side in the inequality~(\ref{Dav}) represents
something more than just an upper bound. It was demonstrated by \citet{Baluev08a}, that in
the LS periodogram case the inequality~(\ref{Dav}) appears rather sharp, especially for
practically important low $\FAP$ levels. In addition to the bound~(\ref{Dav}), we will
deal with the following approximation:
\begin{eqnarray}
\FAP_{\rm max}(Z,f_{\rm max}) = 1 - P_{\rm max}(Z,f_{\rm max}),\nonumber\\
P_{\rm max}(Z,f_{\rm max}) \approx e^{-\tau(Z,f_{\rm max})} P_{\rm single}(Z).
\label{alias-free}
\end{eqnarray}
As it was discussed in \citep{Baluev08a}, the formulae~(\ref{alias-free}) should provide a
good approximation to $\FAP_{\rm max}$ \emph{uniformly} (i.e., for all $\FAP$ levels) in
the case of small aliasing. Note that the approximation~(\ref{alias-free}) and the
bound~(\ref{Dav}) yield almost coinciding results if $\FAP<0.1$, so that the mentioned
property of the approximation~(\ref{alias-free}) probably will not have direct practical
application. In this paper, we use~(\ref{alias-free}) just to plot a reference
`alias-free' $\FAP$ curve.

Now we need to obtain the function $\tau(z,f_{\rm max})$ for our special case of the
multi-harmonic periodograms. In particular, we need to calculate the factor $A(f_{\rm
max})$, present in the expressions for $\tau$. In general, this factor depends in a rather
unpleasant way on the models of the data, on the time series sampling, and on the sequence
of the statistical weights of observations. To attain some technical simplicity, let us
firstly assume that, like in the classical LS periodogram, the base model is empty:
$d_{\mathcal H}=0$ and $\mu_{\mathcal H}(t)\equiv 0$. In this case, we need to find
firstly the eigenvalues $\lambda_k$ of the $d\times d$ matrix $\mathbfss M$, which is
defined as:
\begin{eqnarray}
\mathbfss Q = \overline{\bvarphi \otimes \bvarphi},\qquad \mathbfss S =
\overline{\bvarphi \otimes \bvarphi'_f}, \nonumber\\
\mathbfss R = \overline{\bvarphi'_f \otimes \bvarphi'_f}, \qquad \mathbfss M = \mathbfss
Q^{-1} (\mathbfss R - \mathbfss S^T \mathbfss Q^{-1} \mathbfss S).
\label{QSRM}
\end{eqnarray}
Here the overline denotes the weighted averaging over the time series and the binary
operation $\otimes$ is the dyadic product of vectors ($\bmath x \otimes \bmath y = \bmath
x \bmath y^{\rm T}$), see Appendix~A in \citep{Baluev08a}. The notation $\bvarphi'_f$
stands for the partial derivative of the vectorial function $\bvarphi(t,f)$ over $f$. We
use the expressions from the paper \citep{Davies87} to calculate the factor $A(f_{\rm
max})$. We need to combine eqs.~(3.2,3.3) and the unnumbered equation following after the
eq.~(3.4) from \citep{Davies87} to obtain the formula~(7) in the paper \citep{Baluev08a}
with
\begin{eqnarray}
 A = \frac{\pi^{n-1}}{\Gamma\left(n+\frac{1}{2}\right)}
  \int\limits_0^{f_{\rm max}} {\rm d}f\int\limits_0^\infty
  \left(1 - \frac{1}{\prod_{k=1}^{2n} \sqrt{1 + x \lambda_k(f)}} \right) \frac{{\rm d}x}{x^{3/2}}.
\label{A-gen}
\end{eqnarray}

We need to obtain some more simple, although possibly approximate, expression for the
factor $A$. To do this, we firstly obtain a suitable approximation to the matrix
$\mathbfss M$ and hence to its eigenvalues $\lambda_k$. After that, we can substitute the
approximations for $\lambda_k$ to~(\ref{A-gen}), in order to derive the final
approximation to $A(f_{\rm max})$. We give the associated details, as well as an
assessment of the practical precision of the resulting approximation, in the
Appendix~\ref{sec_CalcA}. Here we give only the final result, which seems to be
sufficiently accurate in practice. The matrix $\mathbfss M$ can be approximated by the
following diagonal block form:
\begin{eqnarray}
\mathbfss M \approx \pi T_{\rm eff}^2
\left(\begin{array}{cccc}
\Id_2 & 0          & \ldots & 0 \\
 0    & 2^2 \Id_2  & \ldots & 0 \\
\ldots & \ldots    & \ldots & \ldots \\
 0    & 0          & \ldots & n^2 \Id_2\\
\end{array}\right),
\label{Mapprox}
\end{eqnarray}
where $\Id_2$ is the $2\times 2$ identity matrix, $T_{\rm eff}$ is the effective time-span
($T_{\rm eff} = \sqrt{4\pi\disp t}$, where $\disp t$ is the weighted variance of timings
$t_i$, see \citealt{Baluev08a}). The approximate equality~(\ref{Mapprox}) implies that the
$2n$ eigenvalues required are grouped into $n$ pairs $\lambda_{2k-1}\approx
\lambda_{2k}\approx \pi T_{\rm eff}^2 k^2,\, k=1,2,\ldots,n$. Finally,
\begin{equation}
 A(f_{\rm max}) \approx 2\pi^{n+\frac{1}{2}} \alpha_n W,
\label{A}
\end{equation}
where $W=f_{\rm max} T_{\rm eff}$ and
\begin{equation}
\alpha_n = \frac{2^{n}}{(2n - 1)!!}
 \sum_{k=1}^{n} \frac{(-1)^{n-k} k^{2n + 1}}{(n+k)!(n-k)!}.
\label{alphaN}
\end{equation}
Here the quantity $(2n-1)!!$ represents the product of all odd integers from $(2n-1)$
downto $1$. The numerical values of the constants $\alpha_n$ for a few values of $n$ are
given in Table~\ref{tab_alpha}.

\begin{table}
\caption{
The constants $\alpha_n$ for a few values of $n$.}
\label{tab_alpha}
\begin{tabular}{@{}ccccccc@{}}
\hline
$n$        & $1$ & $2$     & $3$     & $5$     & $8$                  & $15$ \\
$\alpha_n$ & $1$ & $1.556$ & $1.062$ & $0.136$ & $9.921\cdot 10^{-4}$ & $1.037\cdot 10^{-10}$ \\
\hline
\end{tabular}
\end{table}

Therefore, using eqs.~(7,8) from \citep{Baluev08a}, we obtain for the basic multi-harmonic
periodogram $z(f)$
\begin{equation}
\tau \approx W \alpha_n e^{-z} z^{n-\frac{1}{2}},
\label{tau-z}
\end{equation}
and for the associated modified periodograms $z_{1,2,3}(f)$
\begin{equation}
\tau \approx W \alpha_n
\frac{\Gamma(\frac{N_{\mathcal H}}{2})}{\Gamma(\frac{N_{\mathcal K}+1}{2})} \times
  \left\{\begin{array}{@{}l}
    \left( \frac{2z_1}{N_{\mathcal H}} \right)^{n - \frac{1}{2}}
    \left( 1-\frac{2z_1}{N_{\mathcal H}} \right)^{\frac{N_{\mathcal K}-1}{2}}, \\
    \left( \frac{2z_2}{N_{\mathcal K}} \right)^{n - \frac{1}{2}}
    \left( 1+\frac{2z_2}{N_{\mathcal K}} \right)^{-\frac{N_{\mathcal H}}{2}+1}, \\
    \left( 2\sinh\frac{z_3}{N_{\mathcal K}} \right)^{n - \frac{1}{2}}
    e^{-z_3 \left( 1+ \frac{2n-3}{2N_{\mathcal K}} \right)}.
  \end{array}\right.
\label{tau-z123}
\end{equation}

\section{Numerical simulations}
\label{sec_Simul}
Speaking in terms of the statistical tests theory, two kinds of mistakes can be made in
the signal detection problem: the false alarm and the false non-detection. Our primary
goal was to keep the false alarm probability at some \emph{a priori} small levels
$\FAP<\FAP_*$. This is guaranteed by the theoretical inequality~(\ref{Dav}). Now our goal
is to characterize (given the condition of bounded $\FAP$) the detection power, which is
provided by the actual precision of the $\FAP$ estimation. We noted above that the right
hand side in~(\ref{Dav}) is expected to provide some approximation to the $\FAP$, not just
an upper bound. However, the error of this approximation depends on conditions: in the
case when distant periodogram values are weakly correlated, this approximation should be
precise, and in the case when there exist pairs (or more complicated combinations) of
strongly correlated distant periodogram values, this precision decreases
\citep[see][Appendix~B]{Baluev08a}. In practice, the absence of strongly correlated peaks
means that the periodograms are free from aliases.

Since we have been already prevented (at the given probability $\FAP_*$) from false alarms
by the upper character of the Davies bound~(\ref{Dav}), now we are more interested in
precise approximation of detection thresholds (i.e., such critical values $z_*$ that
provide $\FAP(z_*)=\FAP_*$) rather than of the $\FAP$s themselves, because it is the
detection threshold $z_*$ that determine the detection probability. This means that we
should pay major attention to horizontal deviations between the simulated and theoretical
$\FAP$ curves, rather than to vertical ones.

We now proceed to testing the precision of the theoretical approximations obtained above
using Monte Carlo simulations of $\FAP_{\rm max}$, in the same way as in
\citep{Baluev08a}. When the order of the approximating trigonometric polynomial grows, the
volume of necessary calculations increases significantly due to the following reasons:
\begin{enumerate}
\item The calculations of single values of the multi-harmonic periodogram require to solve
higher-dimensional linear least-squares problem (or to orthogonalize higher-dimensional
functional bases).

\item As the simulations have shown, the average density of peaks on the multi-harmonic
periodograms increase roughly as $\mathcal O(n)$. This requires for the calculations to be
performed on a more dense frequency grid, in order to obtain enough accurate values of
periodogram maxima.
\end{enumerate}
Therefore, our abilities in making numerical simulations are severely limited to small $n$
only.

\begin{figure}
\includegraphics[width=84mm]{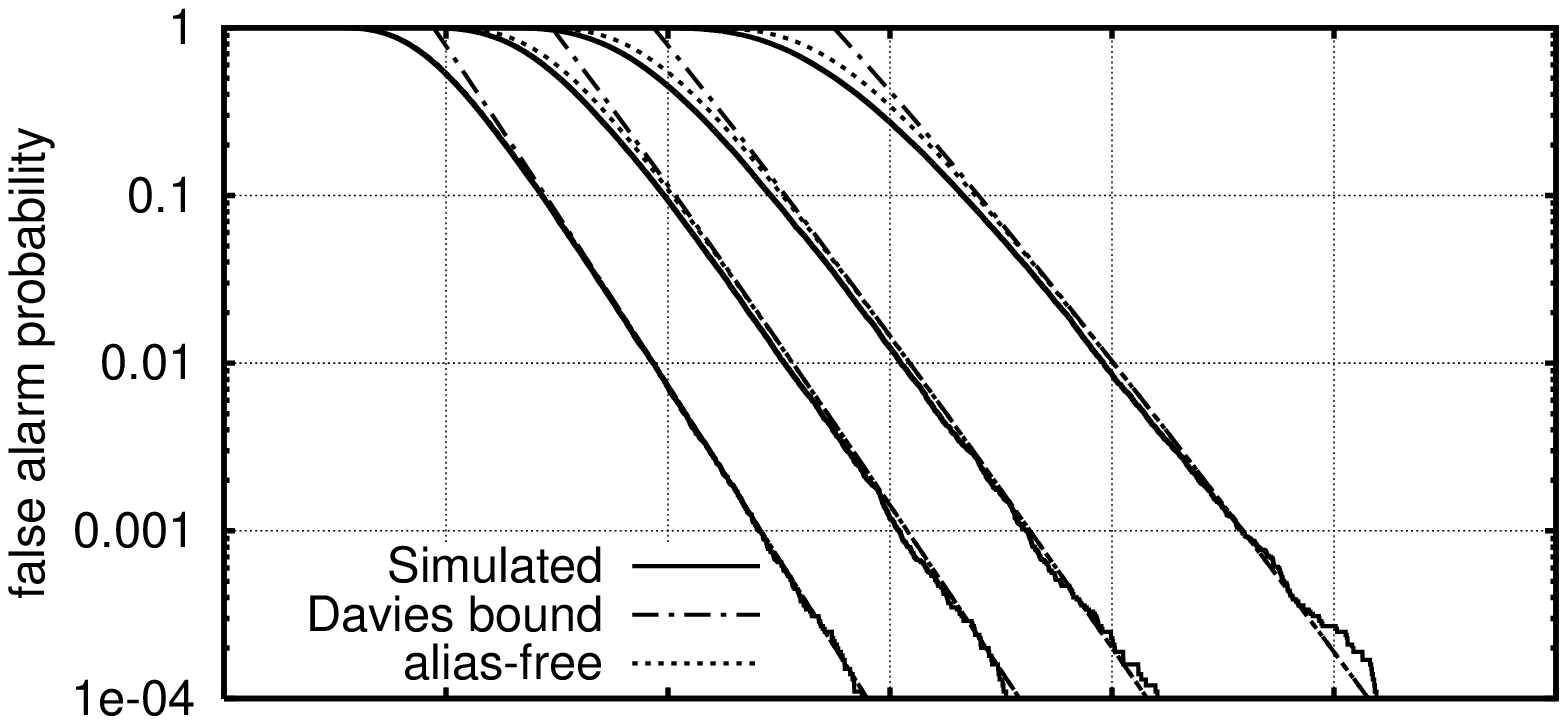}\\
\includegraphics[width=84mm]{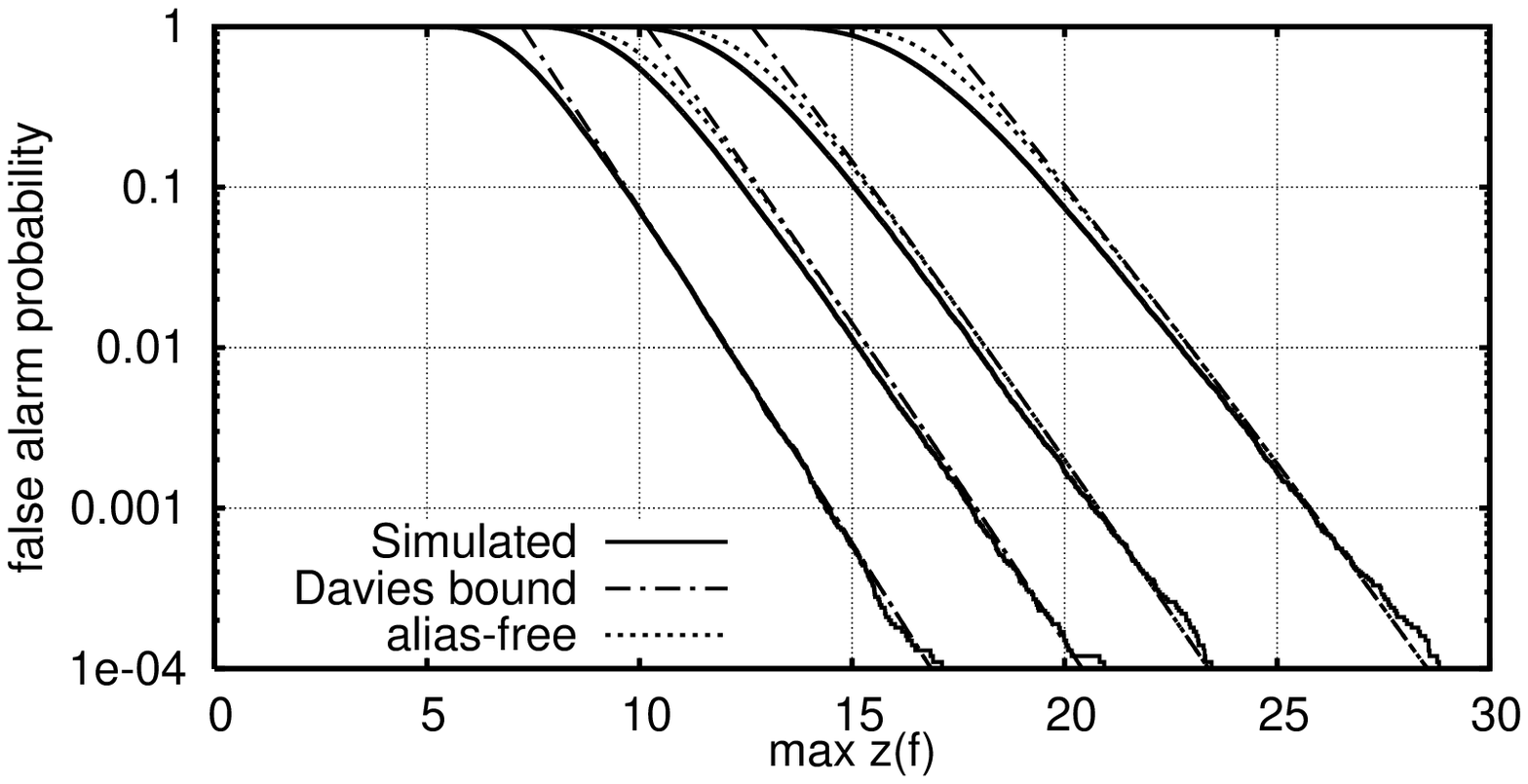}
\caption{Simulated vs. analytic false alarm probability for the multi-harmonic periodogram
of $N=1000$ evenly spaced observations. Results for $n=1,2,3,5$ are shown as converging
bunches of curves from left to right on each panel. The frequency bandwidth was $f_{\rm
max} T = 50$ (top panel) and $500$ (bottom panel). Here and in all other similar figures
further, the number of Monte Carlo trials was about $10^5$ for each simulation curve.}
\label{fig_simeven}
\end{figure}

Firstly let us deal with the case when the time series does not produce any aliasing in
the classical sense, i.e. on the LS periodogram. This is the case of a large number of
evenly distributed observations. The corresponding simulated $\FAP_{\rm max}$ curves are
shown in Fig.~\ref{fig_simeven} for the periodograms $z(f)$. We can see that the
theorectical approximations work quite well. Nevertheless, the small deviations for the
cases $n\geq 2$ contrast with the LS case $n=1$, for which we cannot see any deviation at
all. Probably these small deviations emerged because of an extra correlation of distant
periodogram values, caused by the fact that the model~(\ref{multiharm-model}) incorporates
several sinusoidal harmonics instead of one. Thus the periodogram values at two
frequencies $f_1$ and $f_2$ appear correlated if $f_1/f_2 \approx p/q$ for some integers
$p,q$ not exceeding $n$ (for $n=1$ we had only the trivial condition $f_1\approx f_2$).
Nevertheless, this subtle self-aliasing\footnote{One may argue that such undestanding of
the notion `aliasing' is not traditional, because the associated effect is not connected
with uneven time series sampling, and is only a result of an interplay between the main
period and its subharmonics. Nevertheless, for the sake of a uniform terminology, we name
here all `wrong' periodogram peaks as aliases, and the associated phenomenon of
correlativity of distant periodogram values as aliasing.} effect seems to have negligible
influence on the precision of our analytic $\FAP$ estimation, at least for $n\leq 5$. The
corresponding errors of periodogram detection thresholds are about (or less) $2-3$ per
cent in these cases. Since the signal amplitude scales roughly as $\sqrt z$, this results
in only $\sim 1$ per cent inaccuracy in the amplitude thresholds.

\begin{figure*}
\includegraphics[height=0.1604\textwidth]{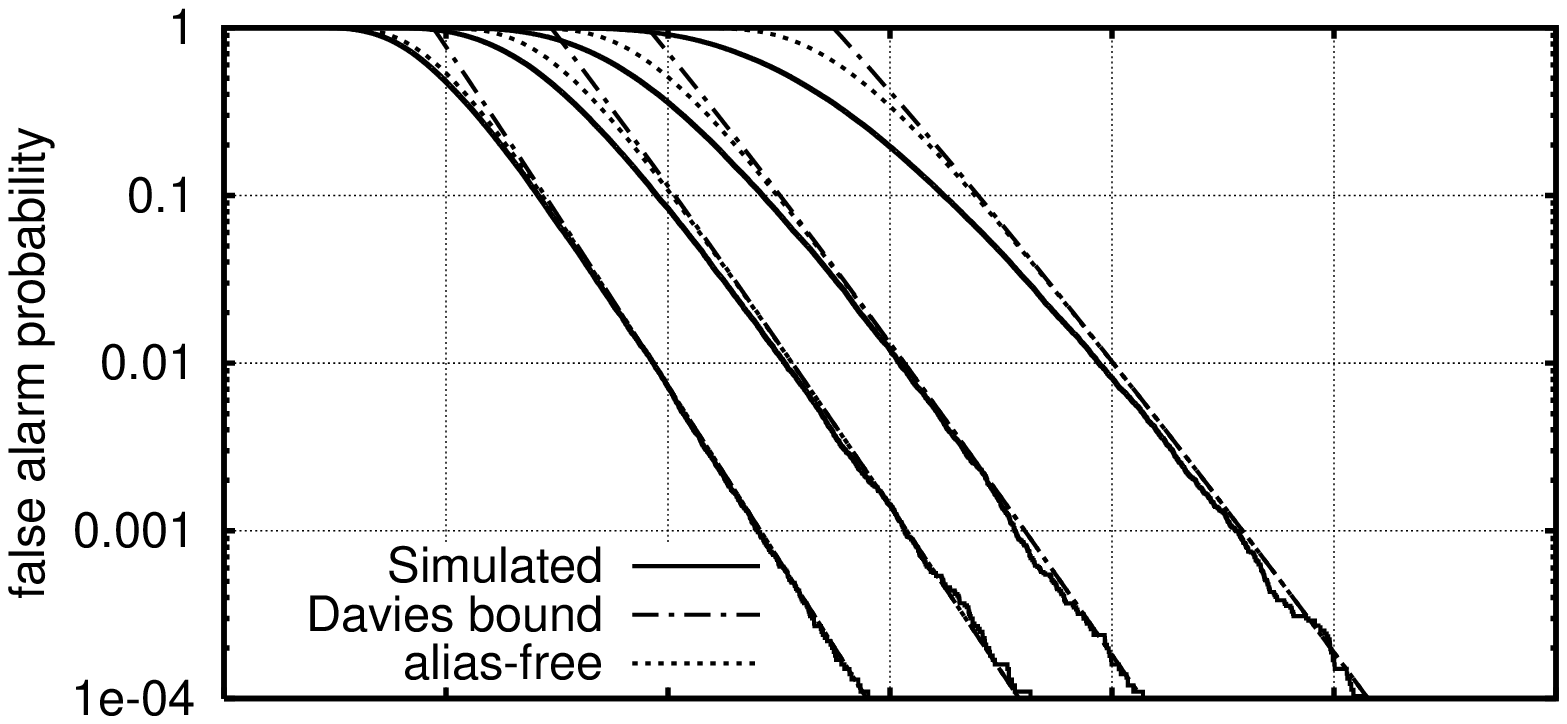}
\includegraphics[height=0.1604\textwidth]{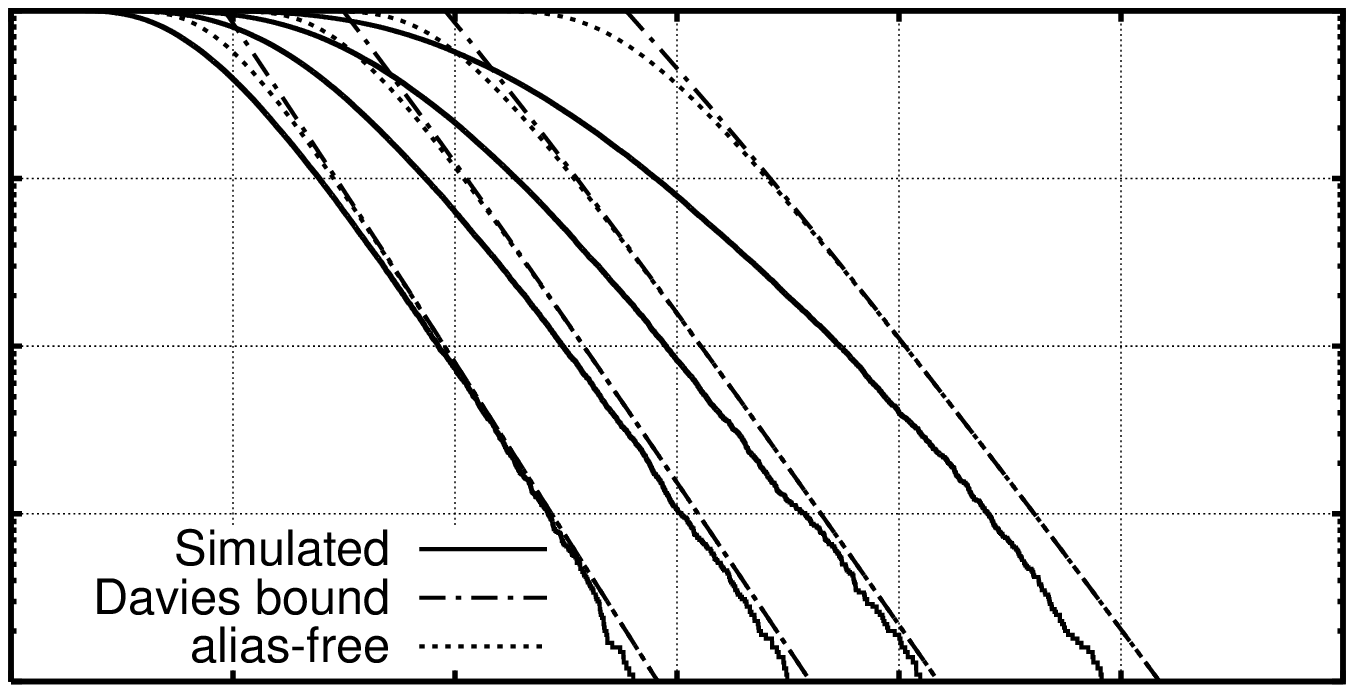}
\includegraphics[height=0.1604\textwidth]{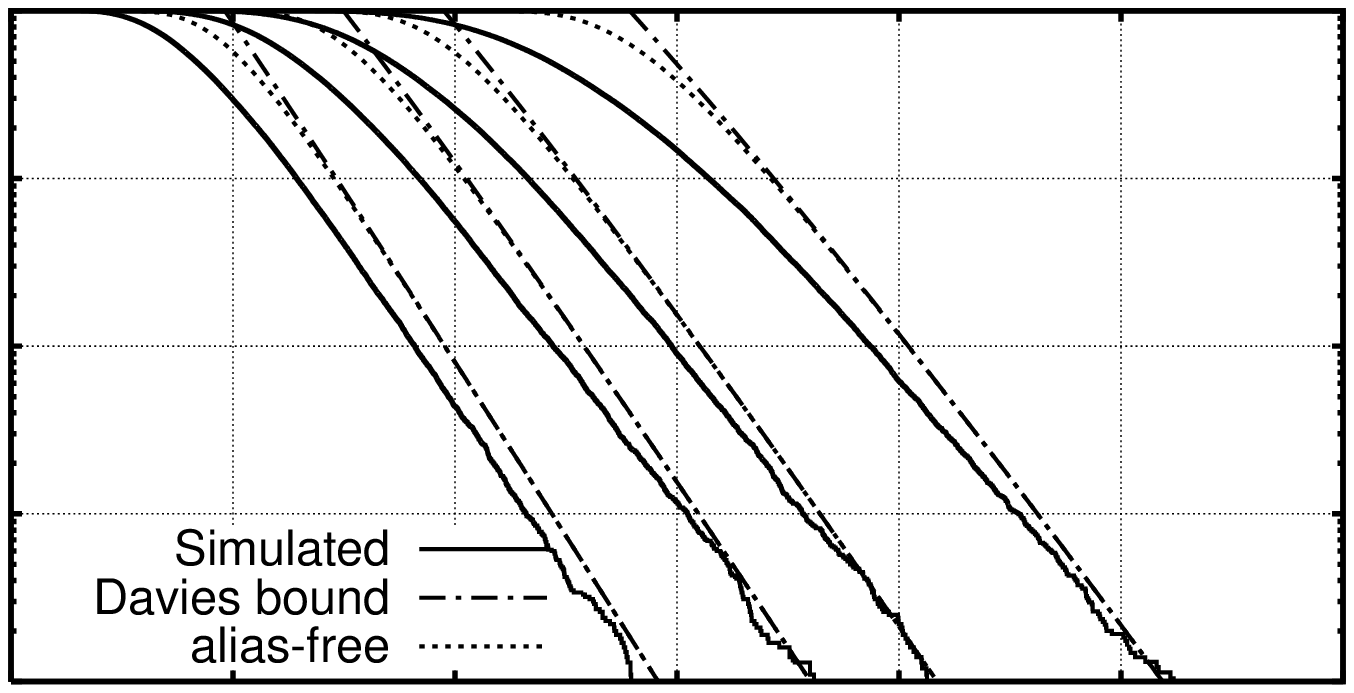}\\
\includegraphics[height=0.1604\textwidth]{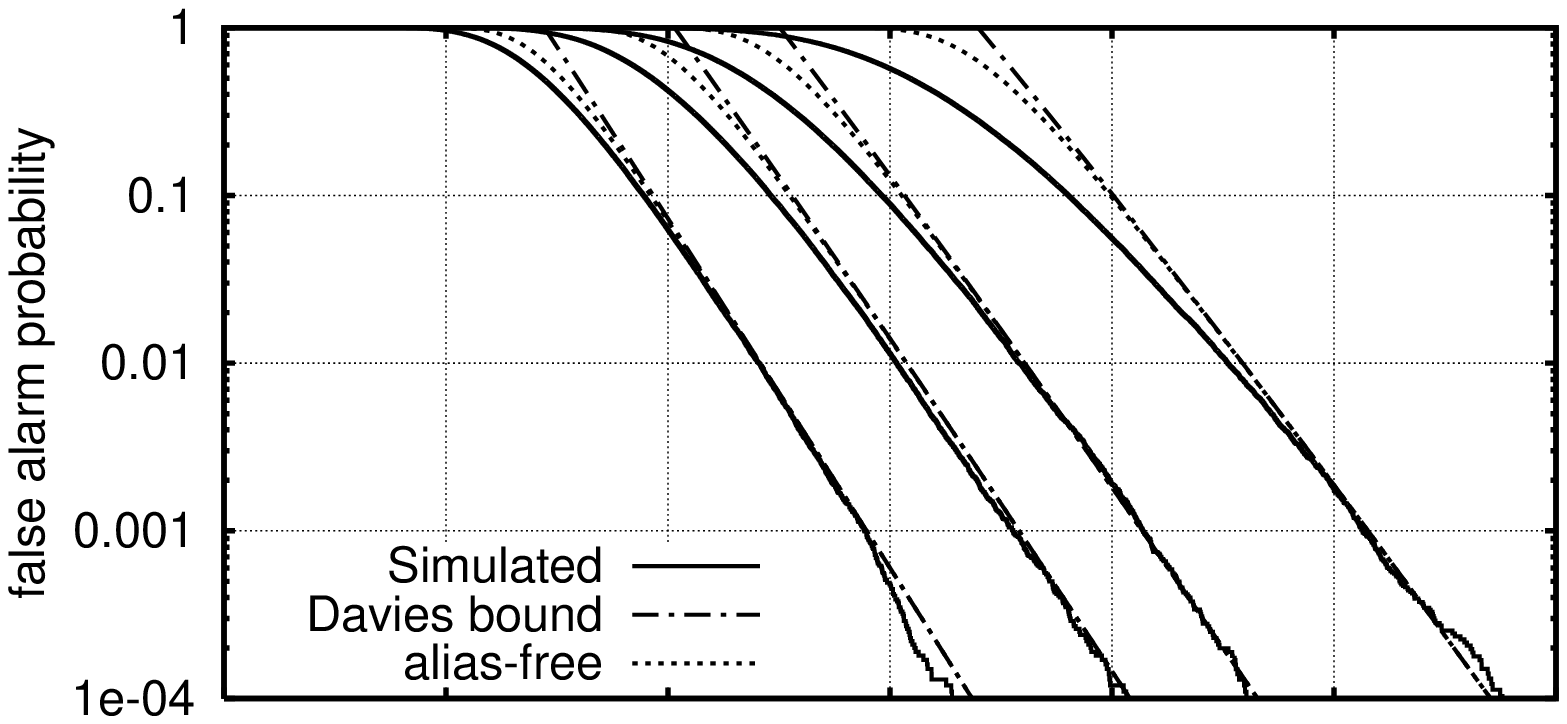}
\includegraphics[height=0.1604\textwidth]{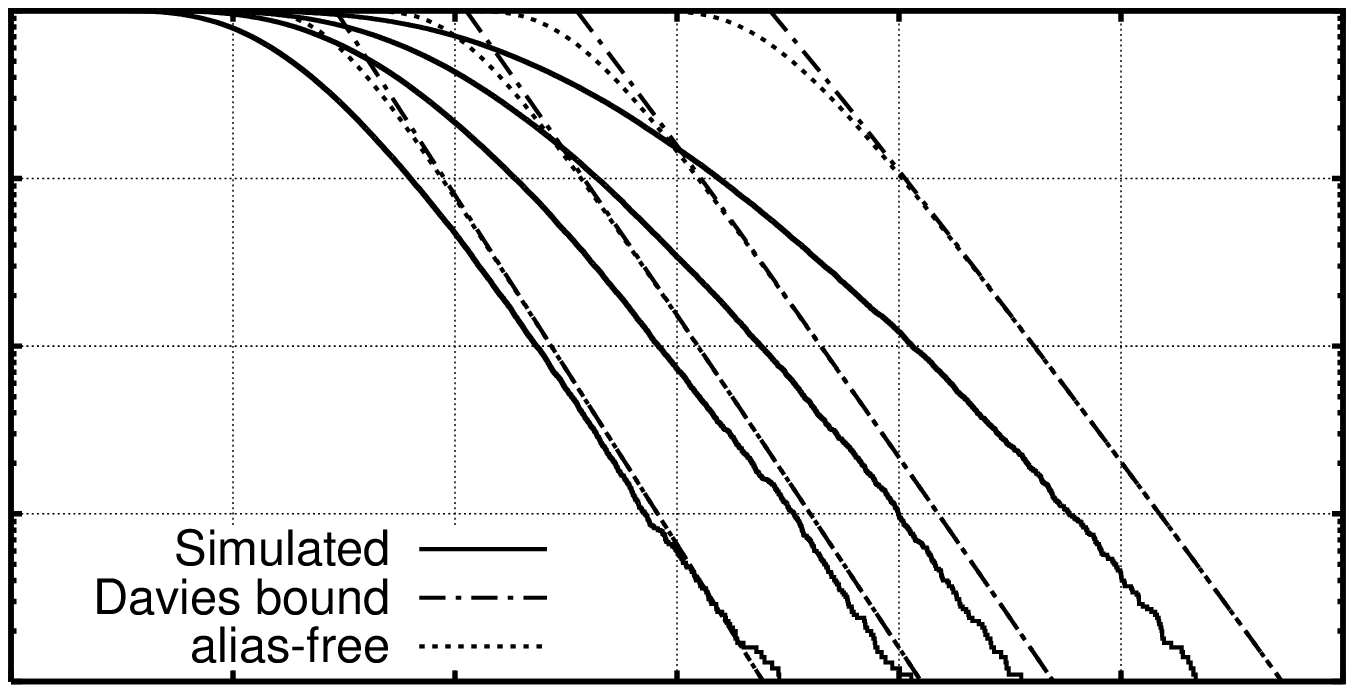}
\includegraphics[height=0.1604\textwidth]{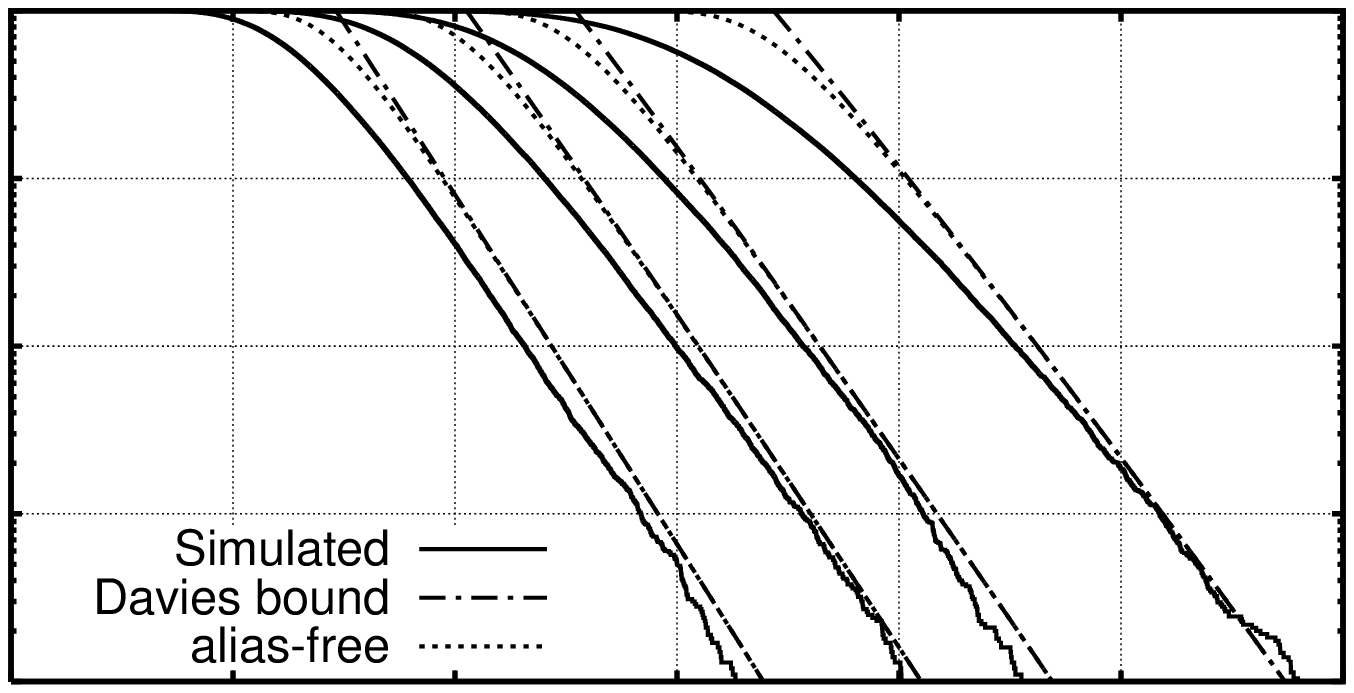}\\
\includegraphics[height=0.1800\textwidth]{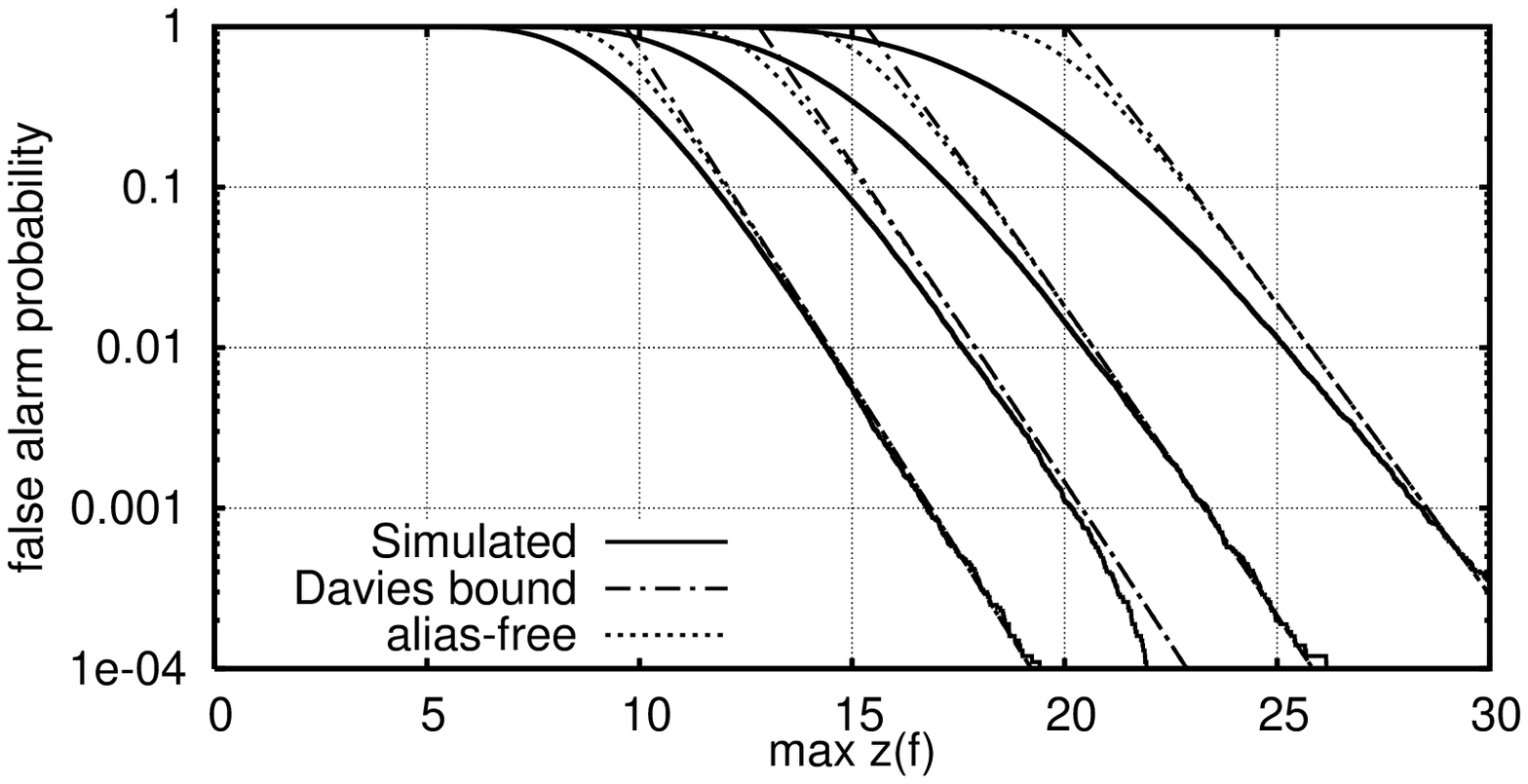}
\includegraphics[height=0.1800\textwidth]{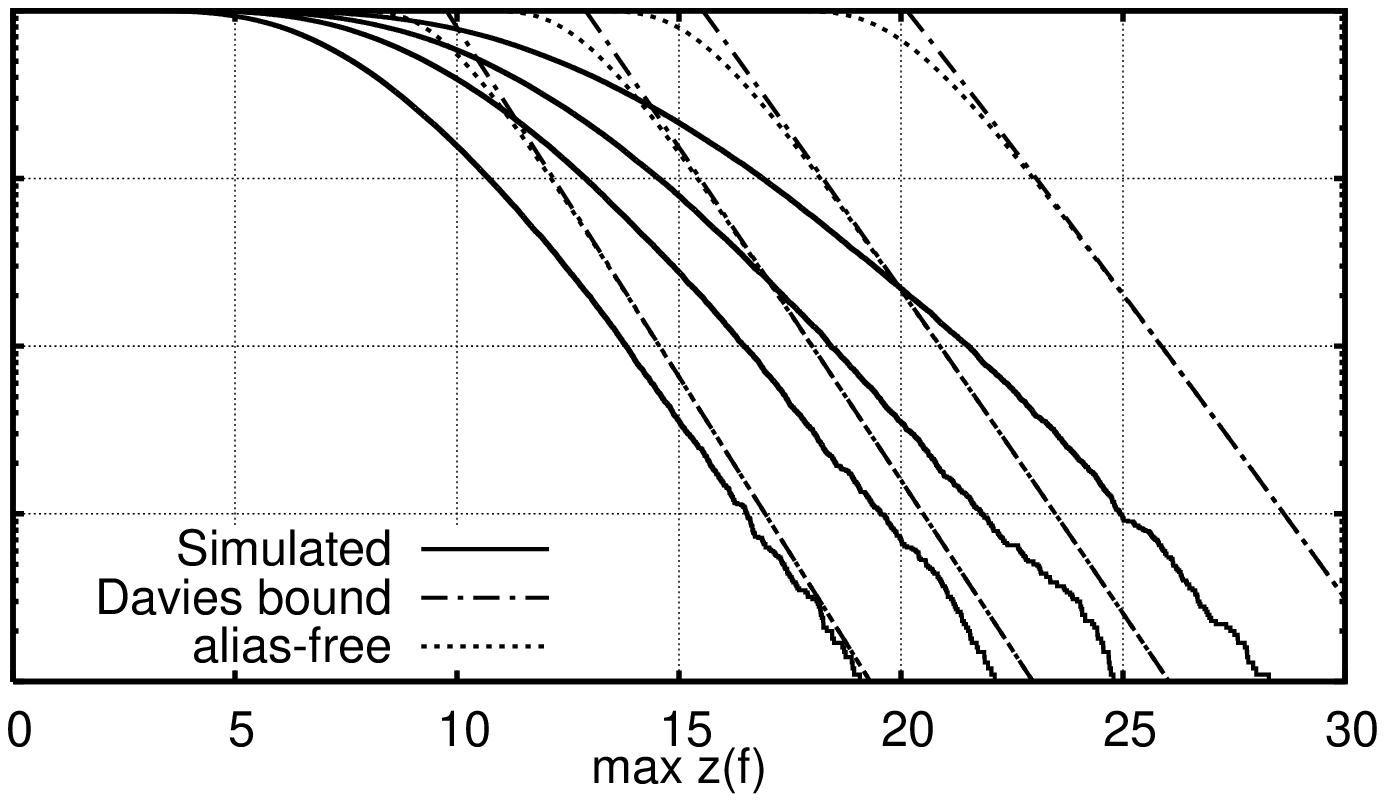}
\includegraphics[height=0.1800\textwidth]{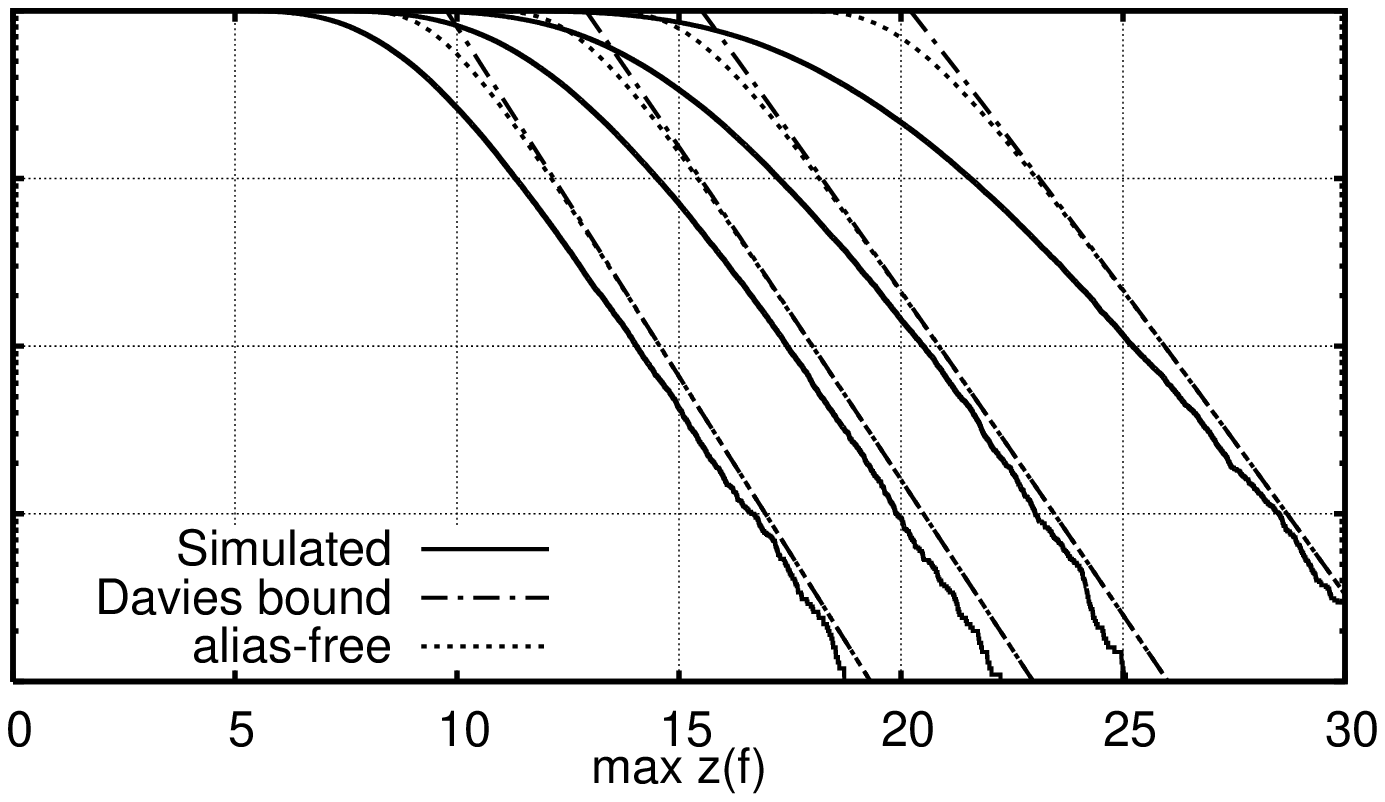}\\
\caption{Simulated vs. analytic false alarm probability for the basic multi-harmonic
periodogram $z(f)$. For the panels in the left and middle columns, the time series
consisted of $N=100$ and $N=30$ randomly spaced datapoints, respectively. For the panels
in the right column $N=100$ datapoints were clumped in ten evenly spaced groups. Each
group consisted of ten points and spanned only $1/50$ fraction of the total time-span
(instead of the natural $1/10$ fraction). In each panel, four converging bunches of curves
from left to right correspond to $n=1,2,3,5$. For the top raw $f_{\rm max}T=50$, for the
middle one $f_{\rm max}T=500$, and for the bottom one $f_{\rm max}T=5000$.}
\label{fig_simrand}
\end{figure*}

When the number of observations decreases and their temporal distribution becomes
non-uniform, the precision of the analytic approximations for $n\geq 2$ decreases in the
same manner as for the usual LS periodogram, $n=1$. Fig.~\ref{fig_simrand} shows a series
of simulations for randomly spaced time series and for a time series with imposed periodic
gapping of timings. We can see that the approximations of the threshold levels $z_*$,
corresponding to $\FAP_*\sim 0.01$, still are rather precise in many cases, which quite
could correspond to a practical situation. The precision of the theoretical approximations
decreases when $f_{\rm max}$ or $n$ grow, when $N$ decreases, or when the degree of the
non-uniformity of timings distribution increases. Nevertheless, even in the worst cases
the relative error of $z_*$ (corresponding to $\FAP_*=0.01$) does not exceed $\sim 20$ per
cent, resulting in only $\sim 10$ per cent overestimation of the corresponding amplitude
thresholds. Such loss of precision still is not catastrophical and quite can be tolerated.

\begin{figure*}
\includegraphics[height=0.1604\textwidth]{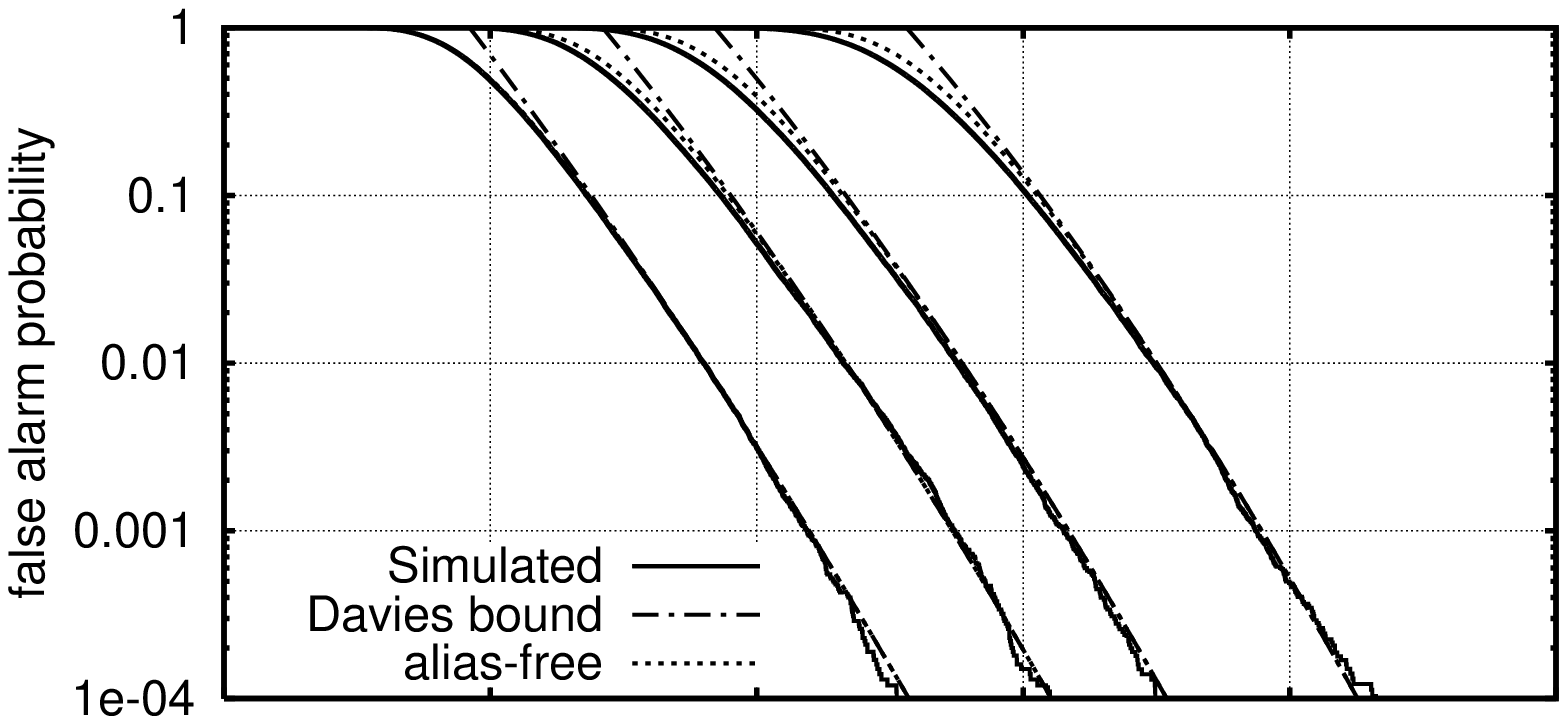}
\includegraphics[height=0.1604\textwidth]{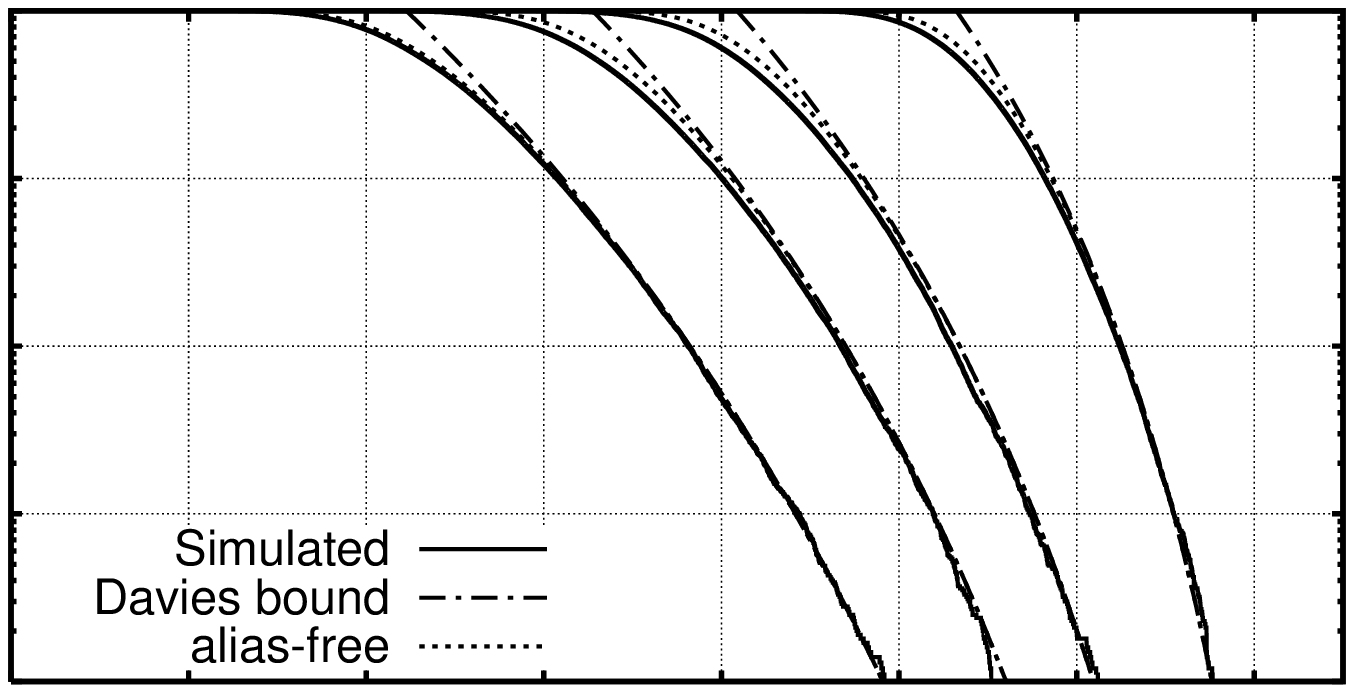}
\includegraphics[height=0.1604\textwidth]{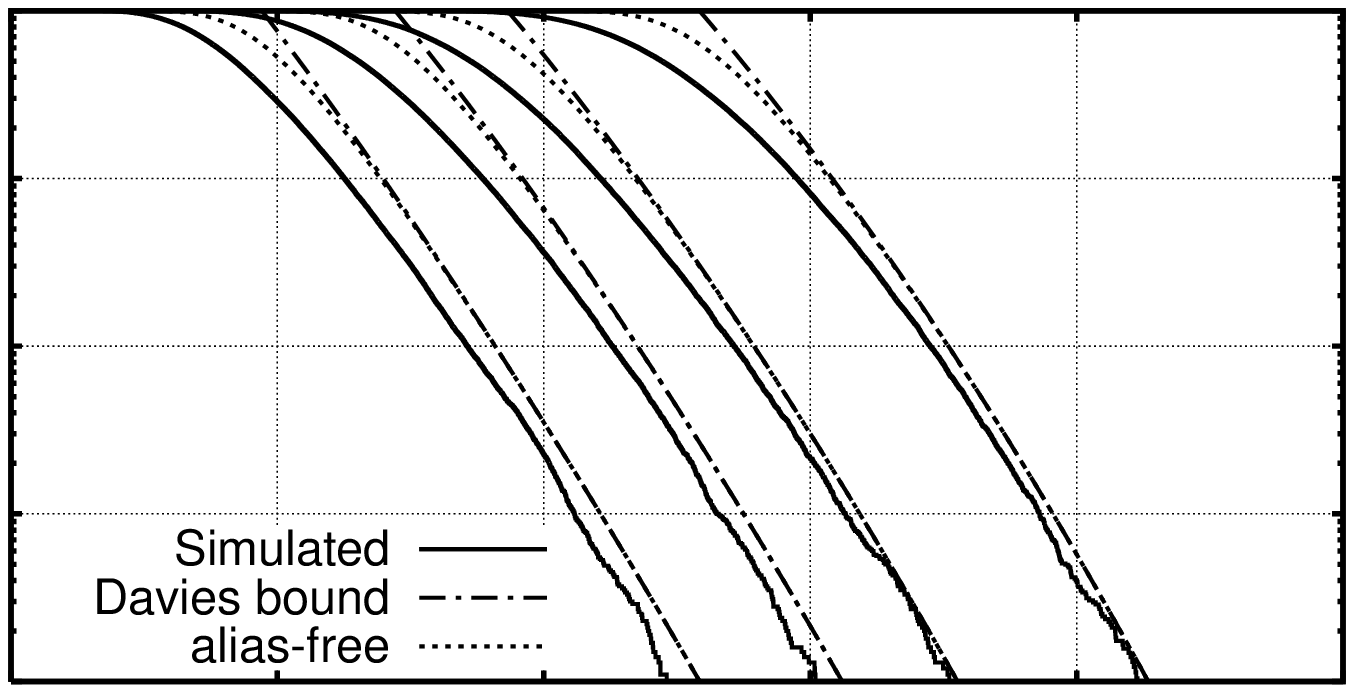}\\
\includegraphics[height=0.1604\textwidth]{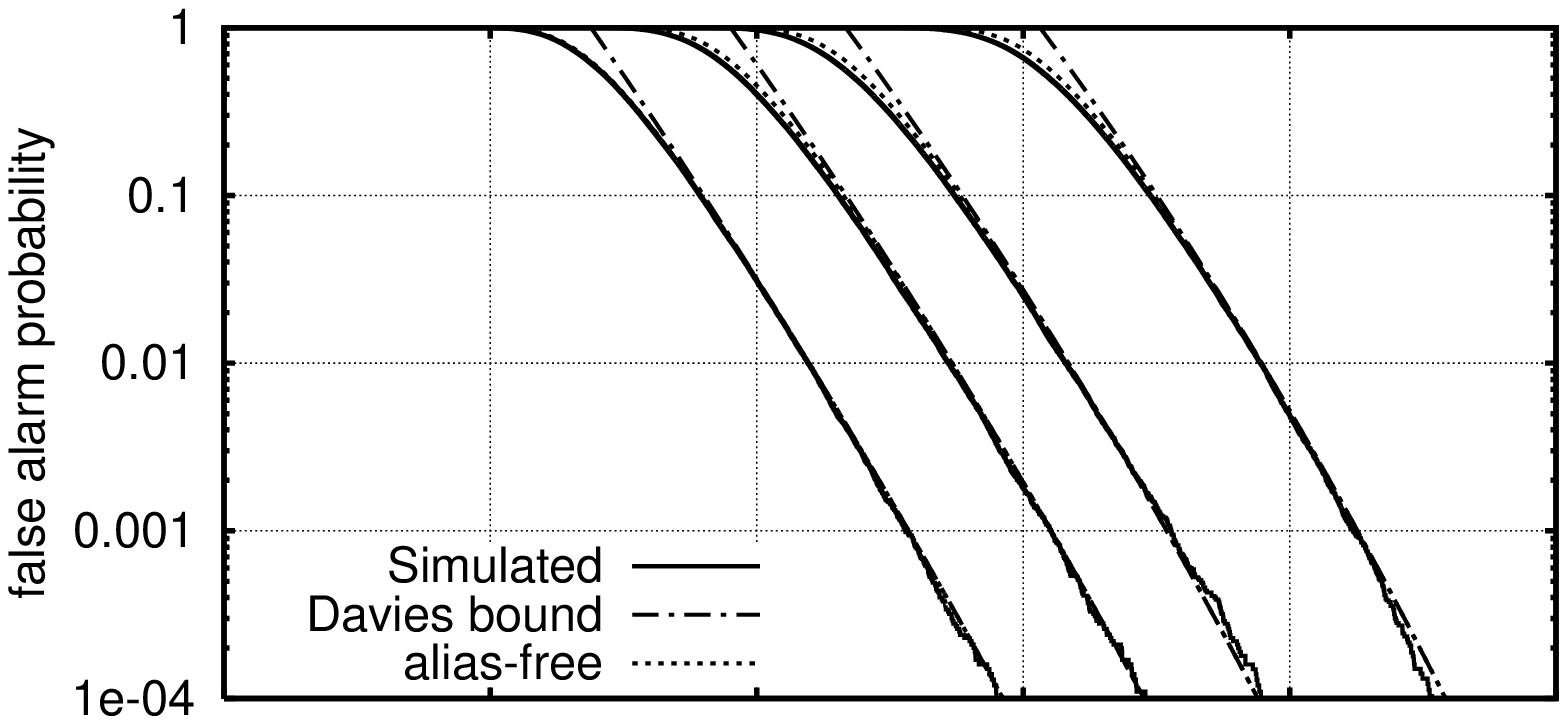}
\includegraphics[height=0.1604\textwidth]{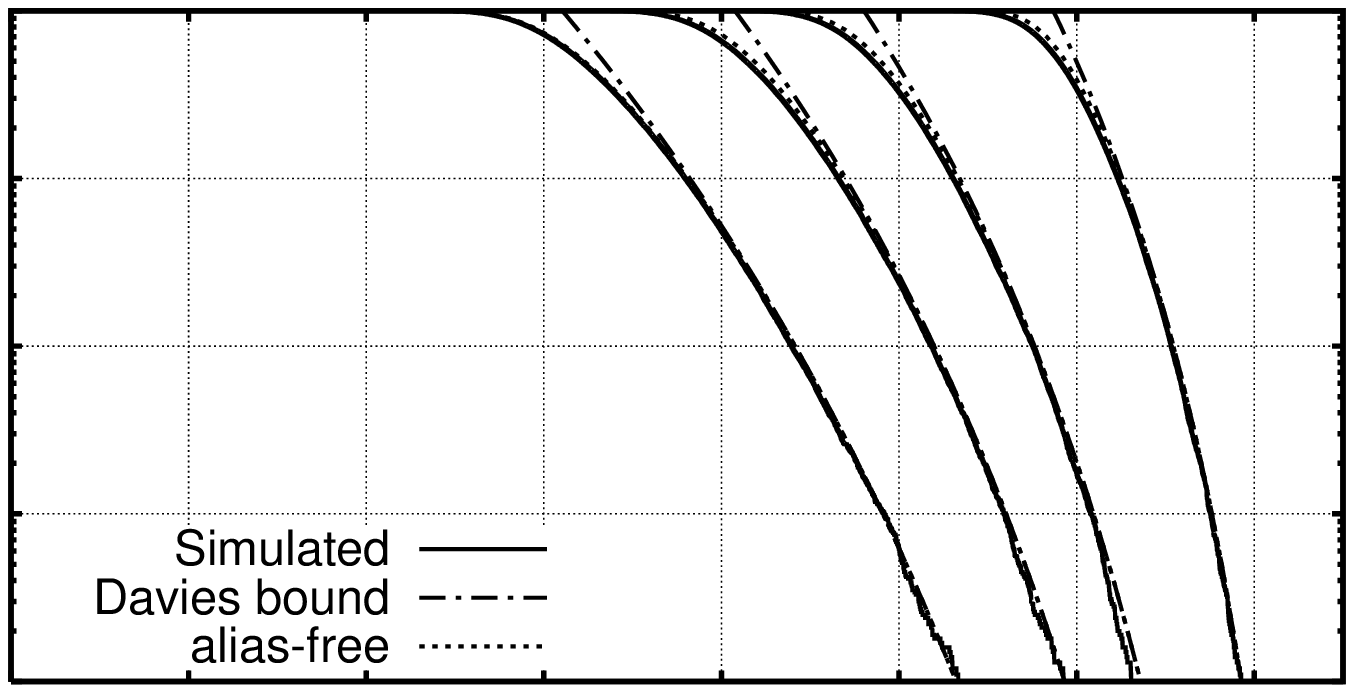}
\includegraphics[height=0.1604\textwidth]{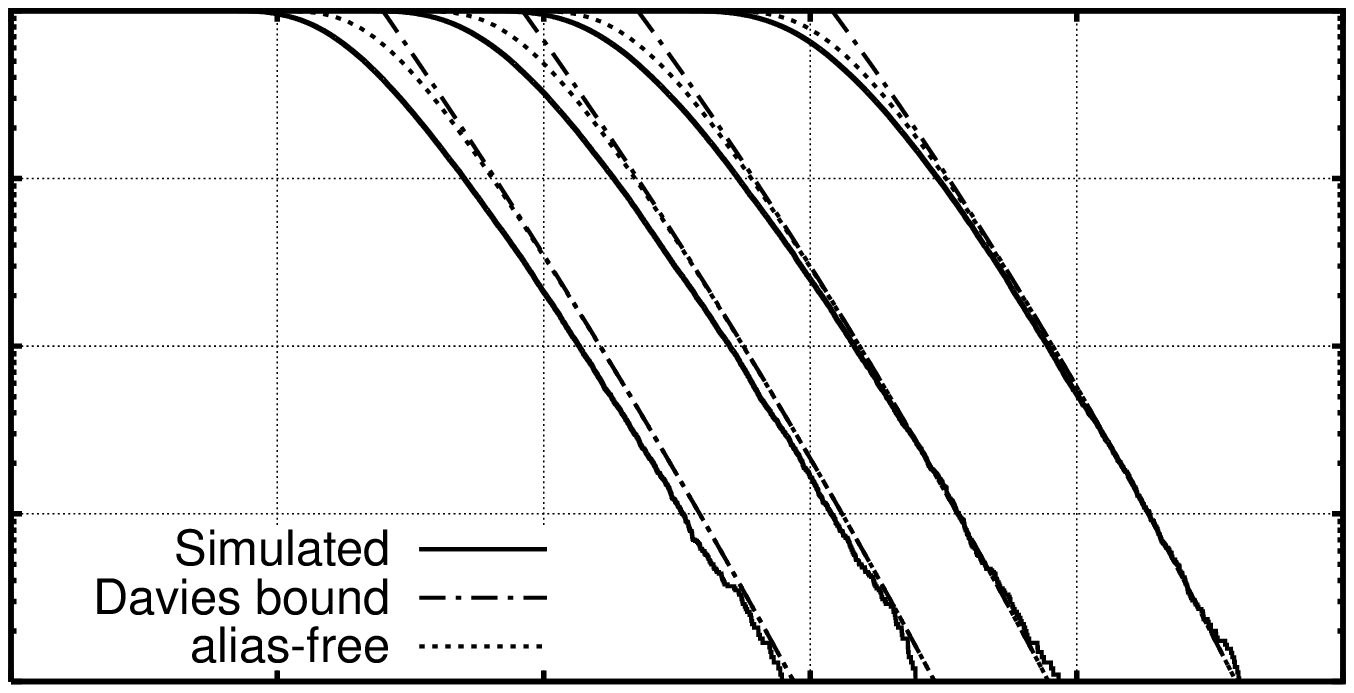}\\
\includegraphics[height=0.1800\textwidth]{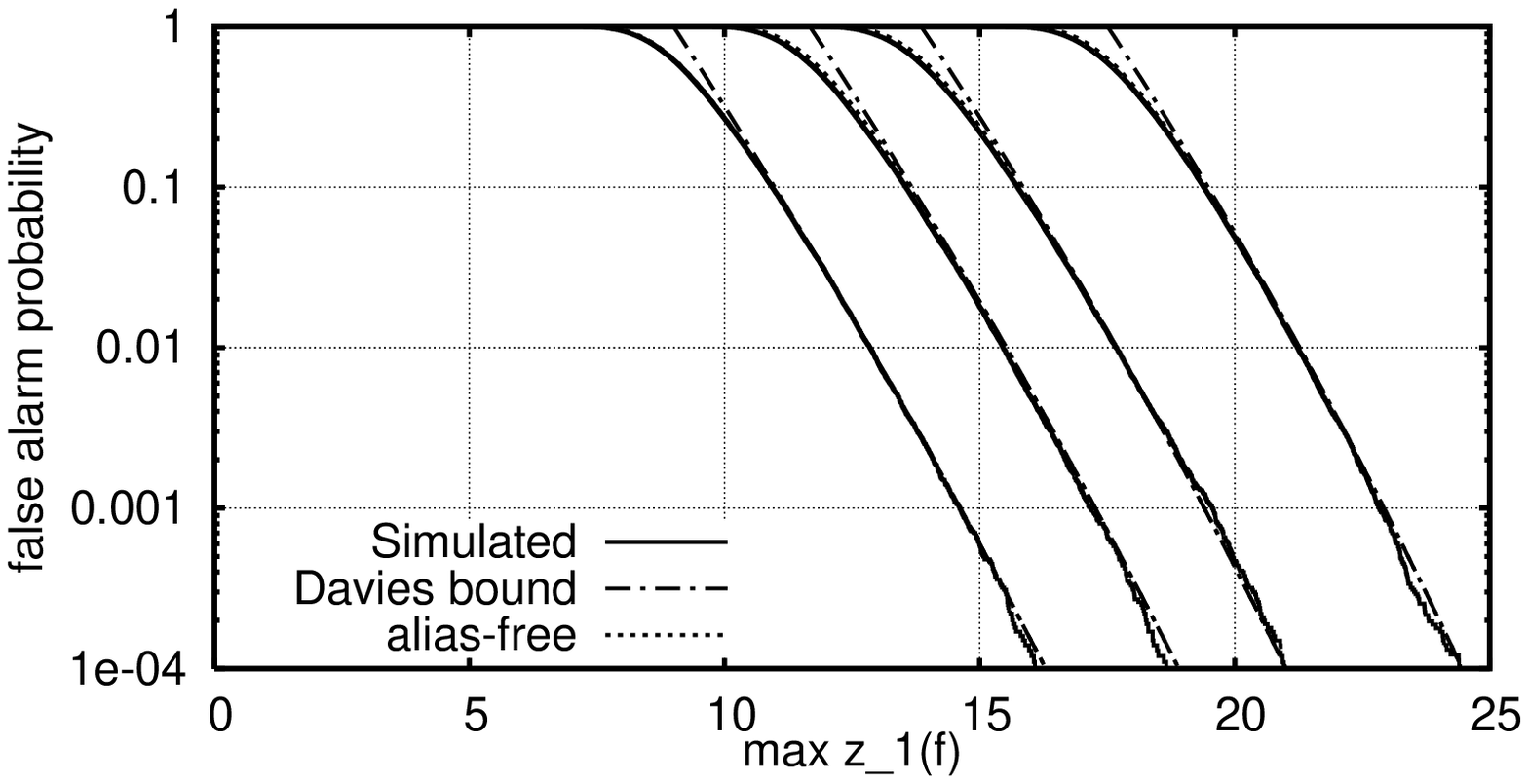}
\includegraphics[height=0.1800\textwidth]{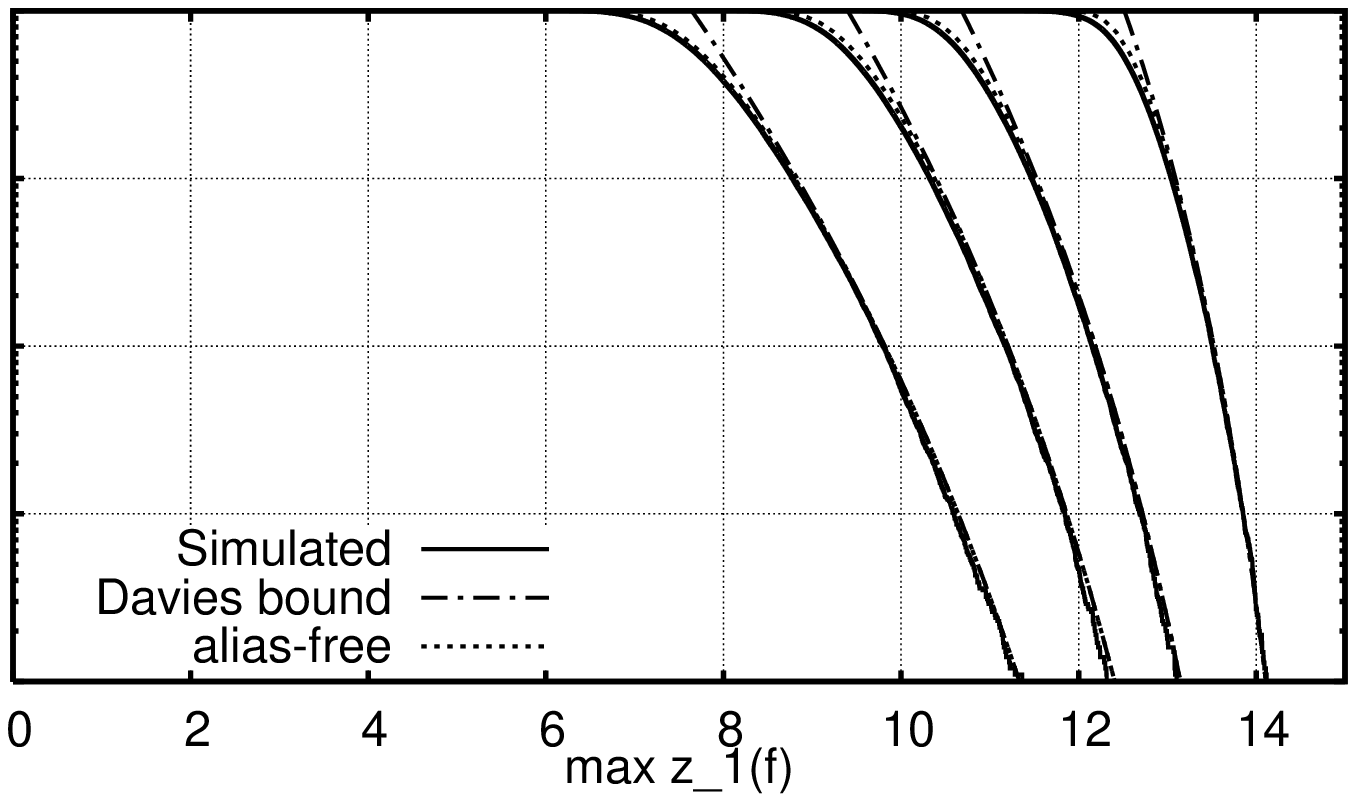}
\includegraphics[height=0.1800\textwidth]{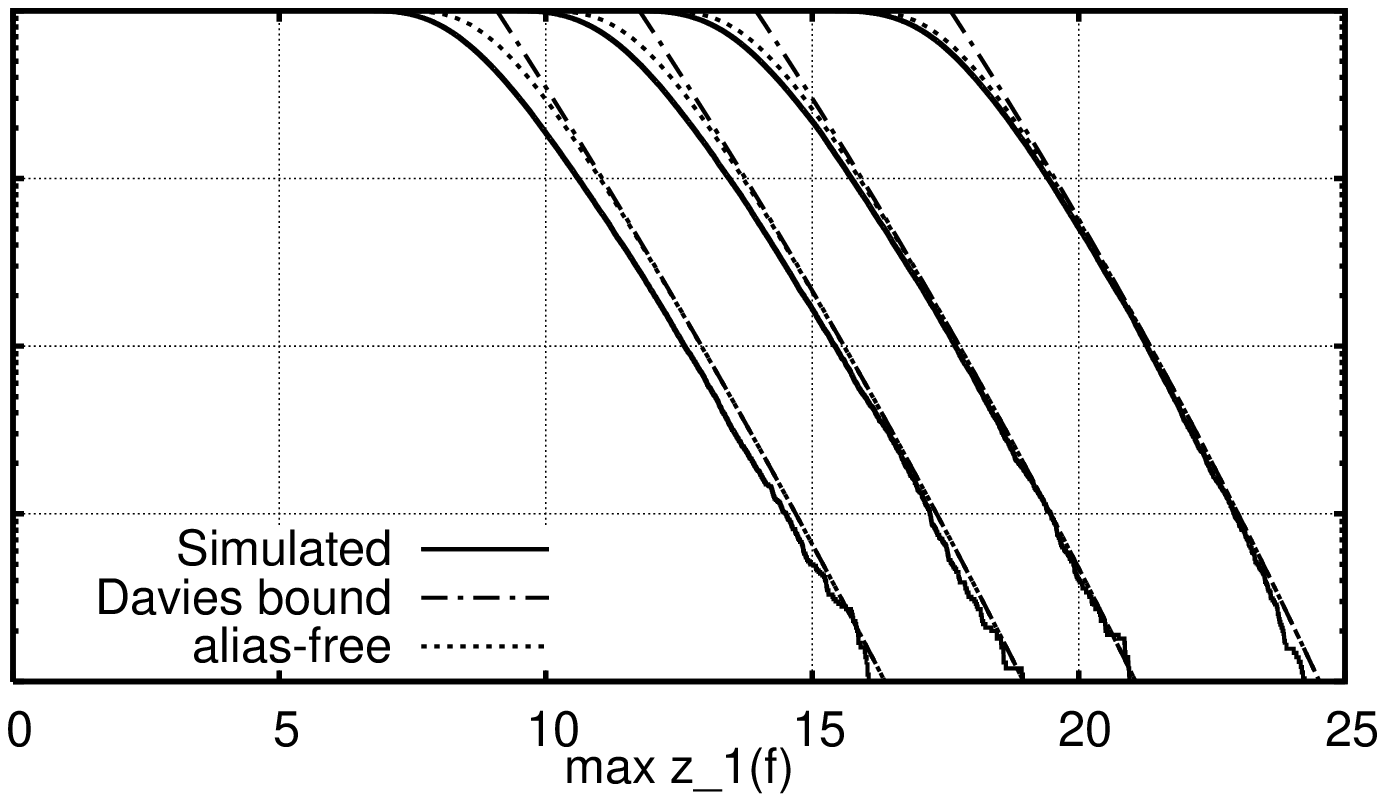}\\
\caption{Same as in Fig.~\ref{fig_simrand}, but for the modified periodogram $z_1(f)$.}
\label{fig_simmod}
\end{figure*}

The paper \citep{Baluev08a} in fact paid undeservedly small attention to the modified LS
periodograms $z_{1,2,3}(f)$. It was assumed that the behaviour of their $\FAP$ curves is
similar to the behaviour of the $\FAP$ curves of the basing LS periodogram. However, they
are the modified periodograms which are usually used in practice. Here we try to correct
this mistake. It appears that the $\FAP$ curves of modified periodograms are considerably
less sensitive to an uneven time series sampling. Consequently, the precision of the
Davies bound~(\ref{Dav}) and of the alias-free approximation~(\ref{alias-free}) appears
significantly better (see Fig.~\ref{fig_simmod}). For the modified multi-harmonic
periodograms, the random distribution of timings does not introduce any significant
perturbation of the $\FAP$ curve even for $N$ as small as $30$. In this case, the $\FAP$
curves for the modified periodograms perfectly agree with the alias-free
approximation~(\ref{alias-free}). For periodically gapped timings, the precision of the
analytic $\FAP$ estimations improves too. Moreover, this precision does not decrease and
even seem to \emph{increase} when the order $n$ or the frequency bandwidth $f_{\rm max}$
grow. The reason for such refinement of the precision of the analytic estimations of the
$\FAP$ for the modified periodograms is unclear.

\begin{figure}
\includegraphics[width=84mm]{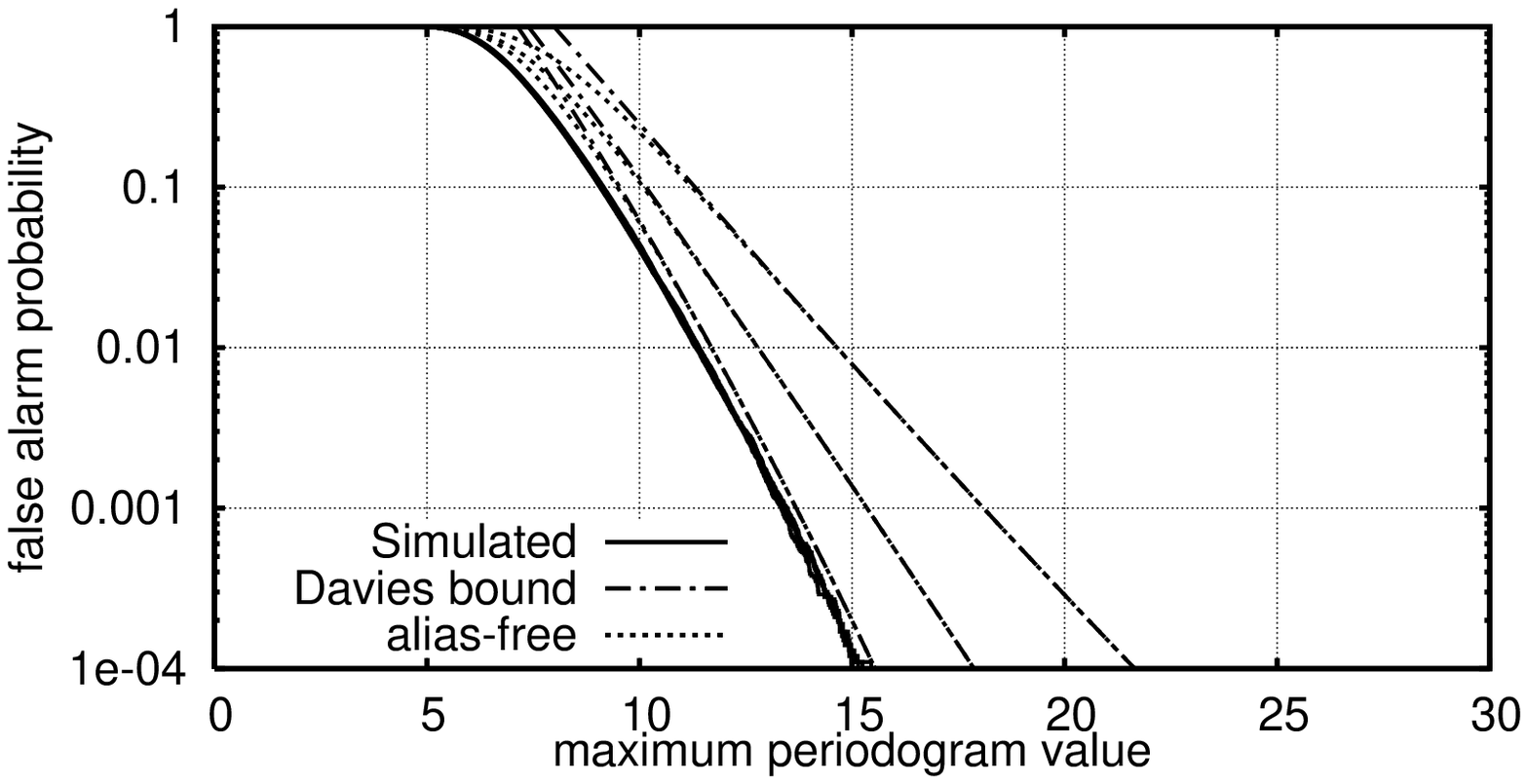}
\caption{Simulated vs. analytic false alarm probability for the modified multi-harmonic
($n=2$) periodogram $z_1(f)$, constructed from $N=100$ evenly spaced observations, in the
main frequency band $f_{\rm max}T=50$. The graph shows \emph{four} simulated $\FAP$ curves
for different degrees of the polynomial trend in the base model $\mu_{\mathcal H}$: empty
base model ($d_{\mathcal H}=0$), a constant term ($d_{\mathcal H}=1$), a linear trend
($d_{\mathcal H}=2$), a quadratic trend ($d_{\mathcal H}=3$). All these curves appear
almost coinciding. For an intercomparison, we show here the theoretical distribution
curves for all the modified periodograms, $z_1$, $z_3$, and $z_2$ (from left to right).
Note that we plot them only for the case $d_{\mathcal H}=0$, because the similar curves
for $d_{\mathcal H}=1,2,3$ did not show any visible deviation.}
\label{fig_simtrend}
\end{figure}

It is harder to complete a similar series of Monte Carlo simulations for more complicated
cases, e.g. with the base model $\mu_{\mathcal H}$ incorporating at least a constant or a
linear trend. We present only a few examples of such simulations, which nevertheless
further certify the practical efficiency of the closed expresions for the $\FAP$ described
above (Fig.~\ref{fig_simtrend}). Actually, it looks that a low-order polynomial trend in
the base model $\mu_{\mathcal H}$ does not introduce any visible deviation in the
simulated $\FAP$ curve, at least in this particular case.

\begin{figure}
\includegraphics[width=84mm]{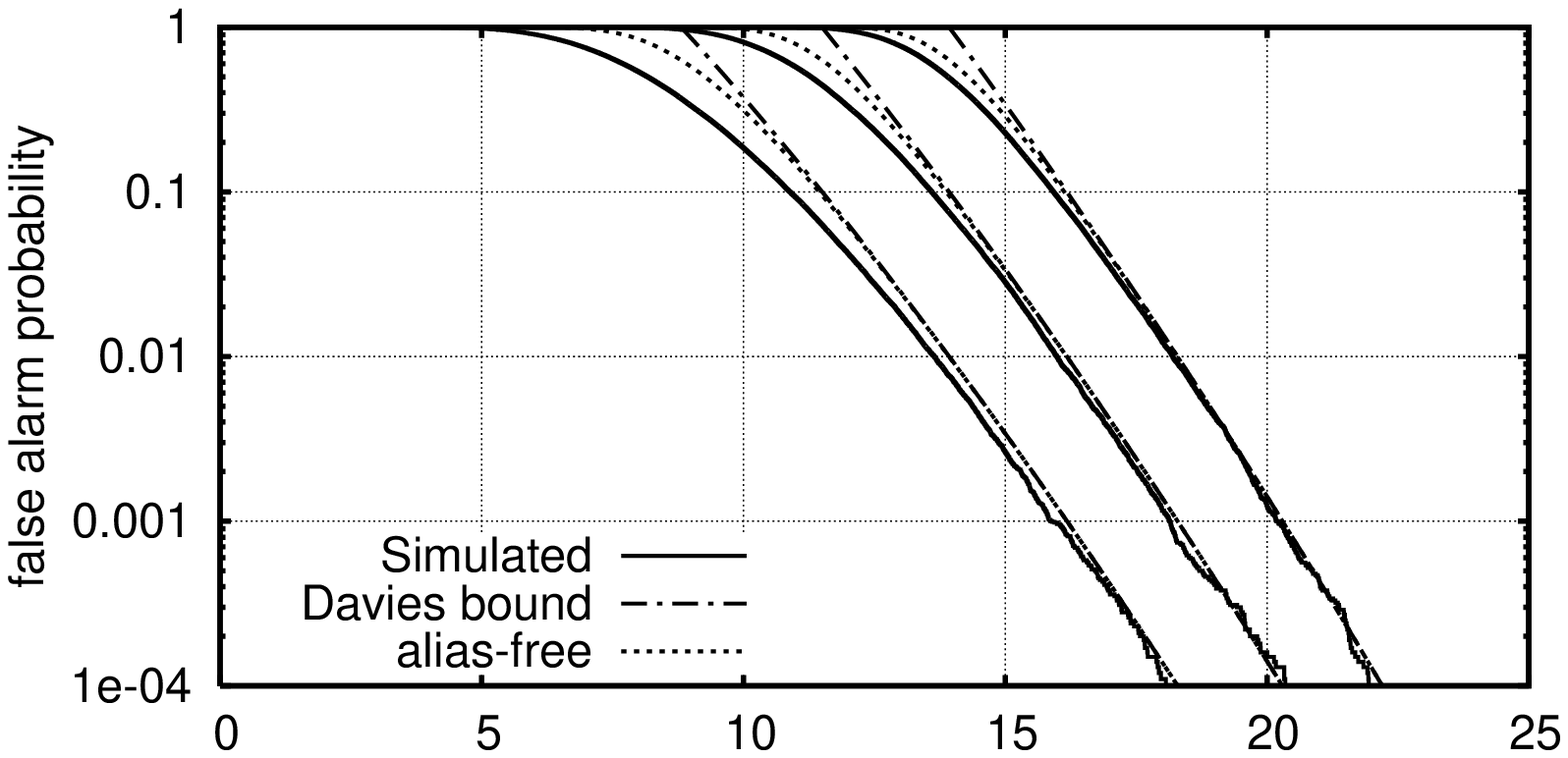}\\
\includegraphics[width=84mm]{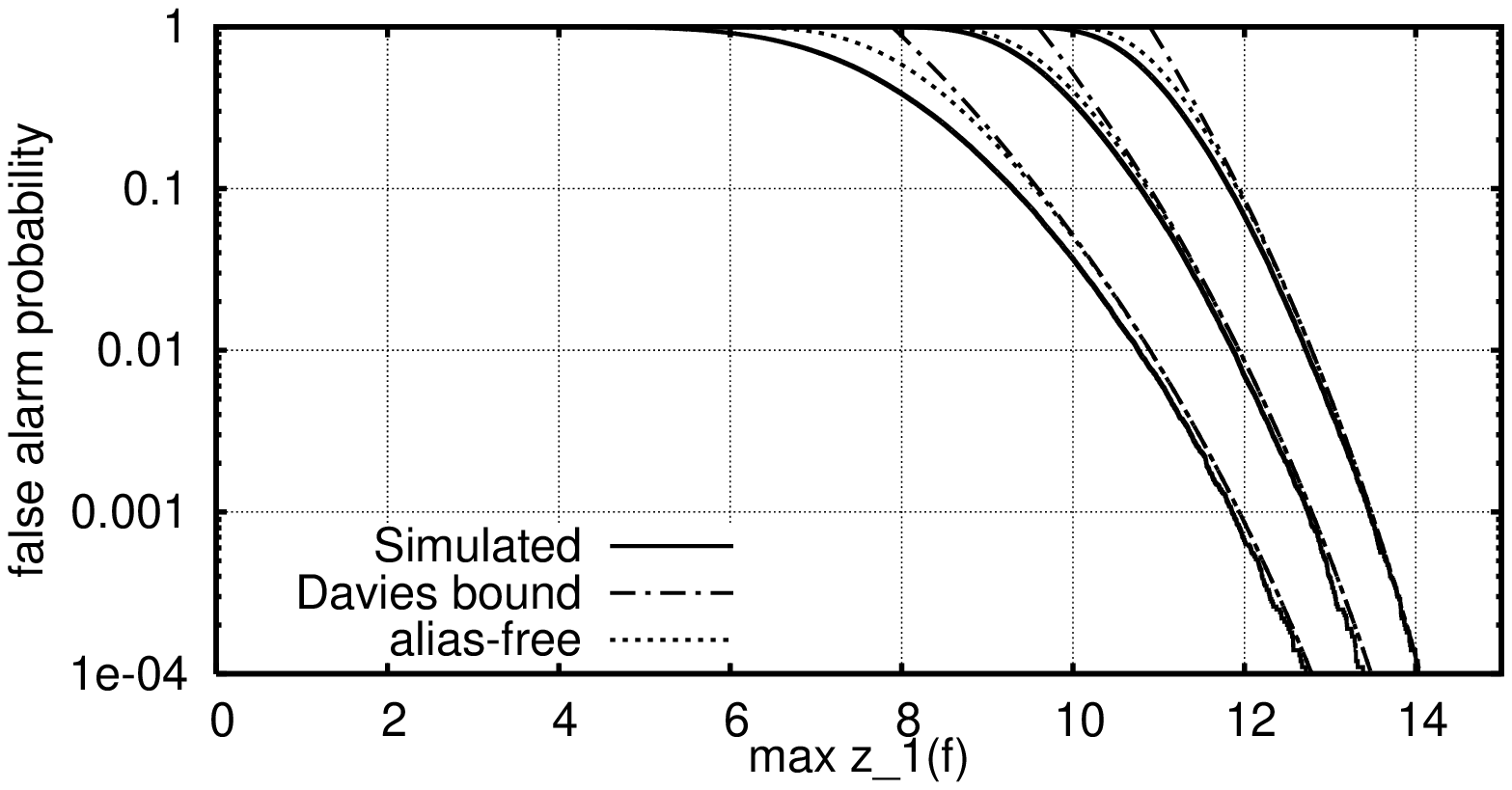}
\caption{Simulated vs. analytic false alarm probability for the multi-harmonic ($n=3$)
periodogram $z_1(f)$ constructed from the radial velocity time series of 51~Peg (top) and
70~Vir (bottom). The base model $\mu_{\mathcal H}$ incorporated a free constant term
($d_{\mathcal H}=1$). In each panel, three bunches of curves correspond to $P_{\rm
min}=1/f_{\rm max}=100$~days, $10$~days, and $1$~day (from left to right).}
\label{fig_sim51Peg70Vir}
\end{figure}

Finally, let us take some realistic time series sampling and consider the associated
$\FAP$ curves and their approximations under some realistic conditions. For this purpose,
as in the paper \citep{Baluev08a}, we use the observational dates and standard errors of
the high-precision radial velocity data for the stars 51~Peg and 70~Vir \citep{Naef04}.
The number of observations in the first time series is $N=153$ and in the second one
$N=35$. The time-span of these time series are about a decade. In both cases, significant
aliasing is present (e.g., corresponding to the annual and diurnal periods). We can see,
however, that in both cases the analytic formulae for the periodogram $z_1$ work very well
(Fig.~\ref{fig_sim51Peg70Vir}).

It is worth noting that in all the cases discussed above, the Davies bound~(\ref{Dav})
indeed bounds the simulated $\FAP$ curves from the upper-right side, at least for not very
small levels $\FAP>10^{-3}$, which can be reliably modelled using $10^5$ Monte Carlo
trials.

\section{Conclusions}
In this paper, previous results by \citet{Baluev08a} are applied to the case when the
model of the signal to be detected represents a truncated Fourier polynomial. Closed
analytic expressions for the false alarm probabilities, associated with multi-harmonic
periodogram peaks, are given. They are tested under various conditions using Monte Carlo
simulations. The simulations have shown that the accuracy of the mentioned theoretical
estimations usually is quite suitable in practice. Also, these simulations have revealed
an unexpected (but pleasant) phenomenon: the accuracy of the above theoretical
approximations of the $\FAP$ is considerably better for the normalized multi-harmonic
periodograms than for the basic, purely least-squares, ones. Since in practice the
observational noise variance is rarely known precisely, they are the normalized
peridograms that are usually dealt with. Therefore, the better behaviour of the $\FAP$
curves for the normalized periodograms has high practical value.

The necessary amount of Monte Carlo simulations for the multi-harmonic periodogram is
bigger than for the LS one. It may appear very difficult to obtain a sufficiently precise
Monte Carlo estimation of the false alarm probability even in the case of a single time
series. Most likely, for surveys dealing with large numbers of separate time series, it
would be impossible to perform the necessary amount of Monte Carlo simulations. For
example, a single CPU at 2 GHz would complete all Monte Carlo simulations presented above
in a few months only. On contrary, the closed theorectical estimations presented in this
paper do not require any simulations at all, and simultaneously often have a nice
accuracy. This indicates that the mentioned estimations represent a promising practical
tool and may be used in a wide variety of astronomical applications, involving search for
non-sinusoidal periodicities in observational data. The corresponding research fields are
ranged from the studies of variable stars to the studies of extrasolar planetary systems.

\section*{Acknowledgments}
This work was supported by the Russian Foundation for Basic Research (Grant 09-02-00230)
and by the Russian President Programme for the State Support of Leading Scientific Schools
(Grant NSh-1323.2008.2). I am grateful to the anonymous referee for providing important
suggestions, which helped to improve the manuscript.

\bibliographystyle{mn2e}
\bibliography{multiharm}

\appendix

\section{
The factor $A(f_{\rm max})$}
\label{sec_CalcA}
The elements of the matrices $\mathbfss Q, \mathbfss S, \mathbfss R$ can be transformed in
the way similar to eqs.~(10) in \citep{Baluev08a}. The matrix $\mathbfss Q$ will contain
the averages of the kind $\overline{\sin k\omega t}$ and $\overline{\cos k\omega t}$. The
matrix $\mathbfss S$ will contain components $\overline{t\sin k\omega t}$,
$\overline{t\cos k\omega t}$, and also $\bar t$. The matrix $\mathbfss R$ will contain
components of the kind $\overline{t^2\sin k\omega t}$, $\overline{t^2\cos k\omega t}$, and
also $\overline{t^2}$. Here $k=1,2,\ldots,2n$. Therefore, we deal with quantities having
the form
\begin{eqnarray}
\Omega_s(f)=\overline{\sin\omega t}, &\quad& \Omega_c(f)=\overline{\cos\omega t}, \nonumber\\
\Lambda_s(f)=\overline{t\sin\omega t}, &\quad& \Lambda_c(f)=\overline{t\cos\omega t}, \nonumber\\
\Xi_s(f)=\overline{t^2\sin\omega t}, &\quad& \Xi_c(f)=\overline{t^2\cos\omega t},
\label{OmegaLambdaXi}
\end{eqnarray}
and with the similar overtonic quantities, calculated at the frequencies
$2f,3f,\ldots,2nf$. Now our goal is to show that under certain conditions the quantities
$\Omega_{c,s},\Lambda_{c,s},\Xi_{c,s}$ have small magnitude in comparison with the
quantities $1,\bar t,\overline{t^2}$, resp., and thus can be neglected. Let us assume that
at the given frequency $f$ the phases $\omega t_i$ are distributed approximately uniformly
in the segment $[0,2\pi]$. This means that the multipliers $\cos\omega t$ and $\sin\omega
t$ in~(\ref{OmegaLambdaXi}) may be considered as random quantities. Their values are
jumping randomly in the segment $[-1,+1]$, whereas the functions $1$, $t$, and $t^2$ are
varying slowly. Therefore, the mentioned sines and cosines may be treated as random
quantities not correlated with the timings $t_i$. This quasirandom property allows us to
write down approximations like $\overline{\sin\omega t}\sim 1/\sqrt N$ and
$\overline{t\sin\omega t}\approx (\bar t) (\overline{\sin\omega t}) \sim \bar t/\sqrt N$.

Therefore, all the quantities~(\ref{OmegaLambdaXi}) may be expected to be negligible (at
the given frequency $f$) when the values of $\Omega_{c,s}(f)$ are small. It is not hard to
see that $\Omega(f) = \Omega_c(f)+i\Omega_s(f) = \overline{e^{i\omega t}}$ (with $i$ being
the imaginary unit) represents the complex spectral window of the time series and the
square of its module is the usual spectral window. Typically, the spectral window contains
a strong narrow peak at $f=0$ and a series of smaller peaks, corresponding to aliasing
frequencies. Therefore, in the case when the spectral window does not contain any strong
peaks at the frequencies $f,2f,\ldots,2nf$, we can keep in the matrices $\mathbfss Q,
\mathbfss S, \mathbfss R$ only the terms, which do not contain sines or cosines inside the
averaging operation. In this approximation, the matrix $\mathbfss M$ can be calculated
easily. The result is given in~(\ref{Mapprox}), and the eigenvalues required are
approximated as $\lambda_{2k}\approx\lambda_{2k-1}\approx \pi T_{\rm eff}^2 k^2$.

It is not hard to check that when our base model $\mu_{\mathcal H}$ is not empty but
contains a free constant term or a low-order polynomial drift with free coefficients, the
same approximation for the matrix $\mathbfss M$ holds true under similar conditions. In
this case, the base model $\mu_{\mathcal H}$ appears approximately orthogonal to the
signal model $\mu$ in the sense that the cross averages $\overline{\bvarphi_{\mathcal
H}\otimes\bvarphi}$ can be neglected in comparison with the respective elements of the
matrices $\overline{\bvarphi_{\mathcal H}\otimes\bvarphi_{\mathcal H}}$ and
$\overline{\bvarphi\otimes\bvarphi}$.

Coupled with the obtained approximate expressions for the eigenvalues $\lambda_k$, the
eq.~(\ref{A-gen}) yields the eq.~(\ref{A}) with
\begin{eqnarray}
\alpha_n = \frac{1}{2\pi\Gamma\left(n+\frac{1}{2}\right)}
  \int\limits_0^\infty \left(1 - \frac{1}{\prod_{k=1}^{n} (1 + x k^2)} \right)
  \frac{{\rm d}x}{x^{3/2}} = \nonumber\\
  = \frac{2^{n}}{(2n-1)!!}\, \frac{1}{2\pi}
  \int\limits_{-\infty}^{+\infty} \left(1 - \frac{1}{\prod_{k=1}^{n} (1 + x^2 k^2)} \right)
  \frac{{\rm d}x}{x^2}.
\label{alpha-gen}
\end{eqnarray}
We can calculate the integral in~(\ref{alpha-gen}) using the theory of functions of a
complex variable. Denoting the integrand in the last integral in~(\ref{alpha-gen}) as
$f(x)$, we can easily check that $\lim_{x\to\infty} |x f(x)| = 0$ (where $x$ is considered
as a complex variable). This means that we can replace the integration line
$(-\infty,+\infty)$ by a closed contour $\mathcal C_R$, representing a semi-circle of the
radius $R\to\infty$ in the upper complex semiplane. Indeed, the integral over the
semicircle arc decays at least as rapidly as $\sim \pi R f(R) \sim \pi/R \to 0$ when
$R\to\infty$, and the integral over the diameter of the semi-circle, $(-R,R)$ tends to the
integral within $(-\infty,+\infty)$ that we need to compute.

The integrand $f(x)$ can be represented as a ratio of two algebraic polynomials:
$f(x)=P(x)/Q(x)$, where $Q(x)=\prod_{k=1}^{n} (1 + x^2 k^2)$ and $P(x)=(Q(x)-1)/x^2$ (it
is not hard to see that $P(x)$ is indeed a polynomial of degree $2n-2$, because the free
term in $Q(x)$ is unit and hence the denominator $x^2$ is reduced). Therefore, the
integral over $\mathcal C_R$ can be expressed via the sum of residues of $f(x)$ in the
points $x_k=i/k,\, k=1,2,\ldots,n$ (with $i$ being the imaginary unit), which represent
the roots of $Q(x)$ in the upper complex semiplane. That is,
\begin{equation}
\frac{1}{2\pi i}\int\limits_{-\infty}^{+\infty} f(x) {\rm d}x = \frac{1}{2\pi i}
 \lim_{R\to\infty} \int\limits_{\mathcal C_R} f(x) {\rm d}x = \sum_{k=1}^{n} \Res
 f(x_k).
\label{intRes}
\end{equation}
Since the singularities $x_k$ are simple poles, the corresponding residues can be
evaluated as
\begin{eqnarray}
\Res f(x_k) = \frac{P(x_k)}{Q'(x_k)} = \frac{k}{2i\prod_{j=1..n,j\neq k}(1-j^2/k^2)} =\nonumber\\
= \frac{k^{2n-1}}{2i\prod_{j=1..n,j\neq k} (k^2-j^2)}
= \frac{(-1)^{n-k} k^{2n+1}}{i (n+k)! (n-k)!}.
\label{Res}
\end{eqnarray}
The formulae~(\ref{alpha-gen},\ref{intRes},\ref{Res}) yield the final
expression~(\ref{alphaN}).

Note that alternatively we could use the treatment involving ellipsoidal surfaces in
multi-dimensional spaces (see eq.~(B7) by \citealt{Baluev08a}). This way seems to be less
convinient to obtain exact formulae for $\alpha_n$, but nonetheless it yields a simple
upper bound
\begin{equation}
\alpha_n \leq \frac{1}{(n-1)!} \sqrt{\frac{1}{n}\sum_{k=1}^n k^2} = \frac{1}{(n-1)!} \sqrt{\frac{(n+1)(2n+1)}{6}}.
\end{equation}
The comparison of this bound with numerical values from Table~\ref{tab_alpha} shows that
this bound is remarkably sharp.

\begin{figure*}
\begin{tabular}{@{}c@{}c@{}}
\includegraphics[width=0.49\textwidth]{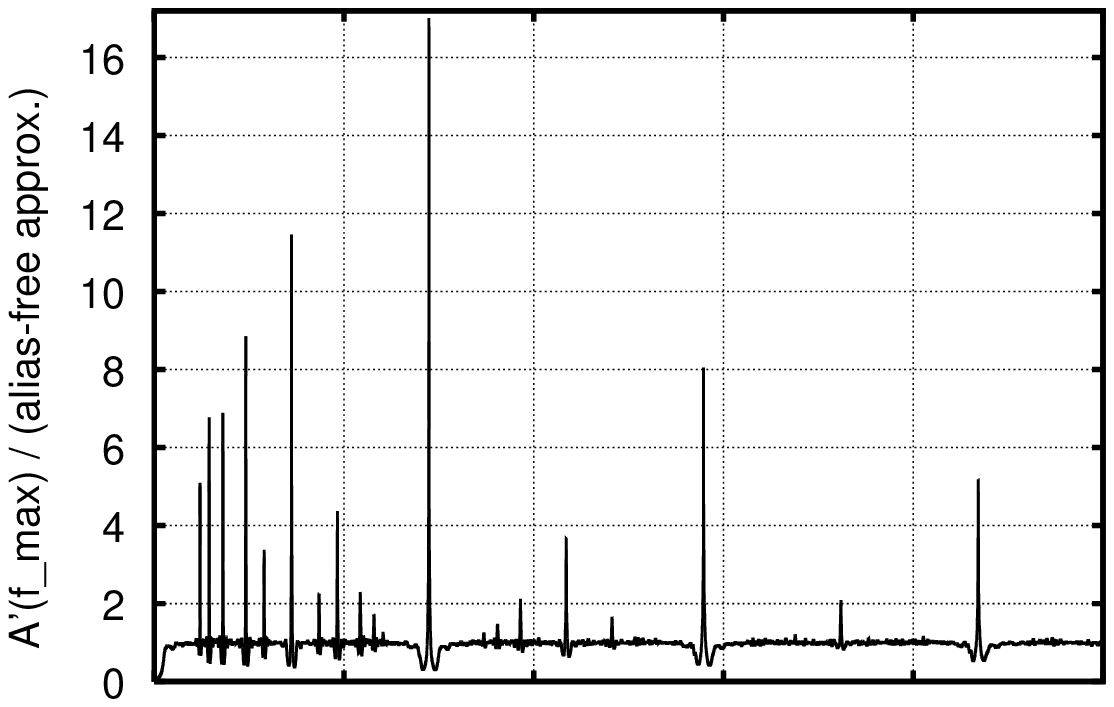} & \includegraphics[width=0.49\textwidth,height=0.30\textwidth]{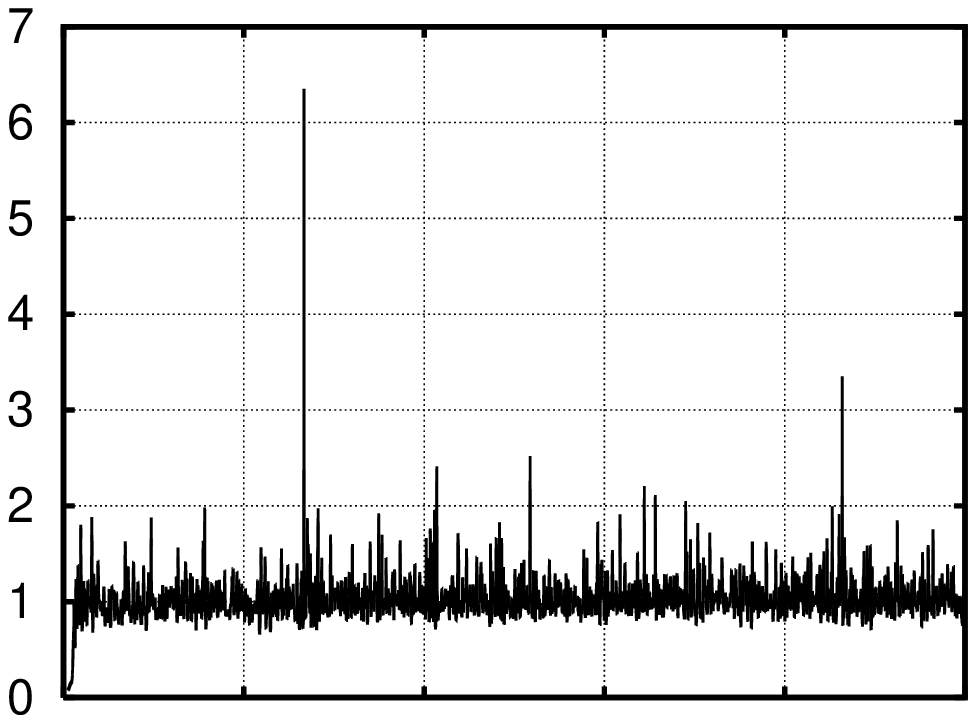} \\
\includegraphics[width=0.49\textwidth]{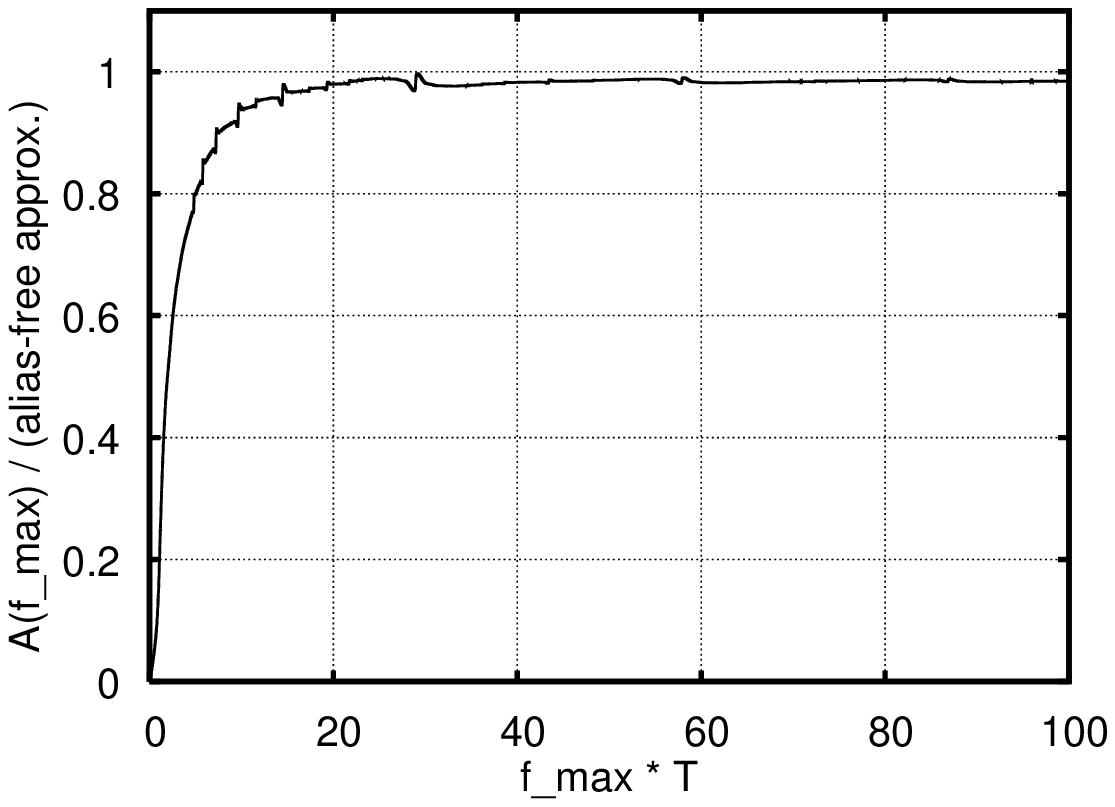} & \includegraphics[width=0.49\textwidth]{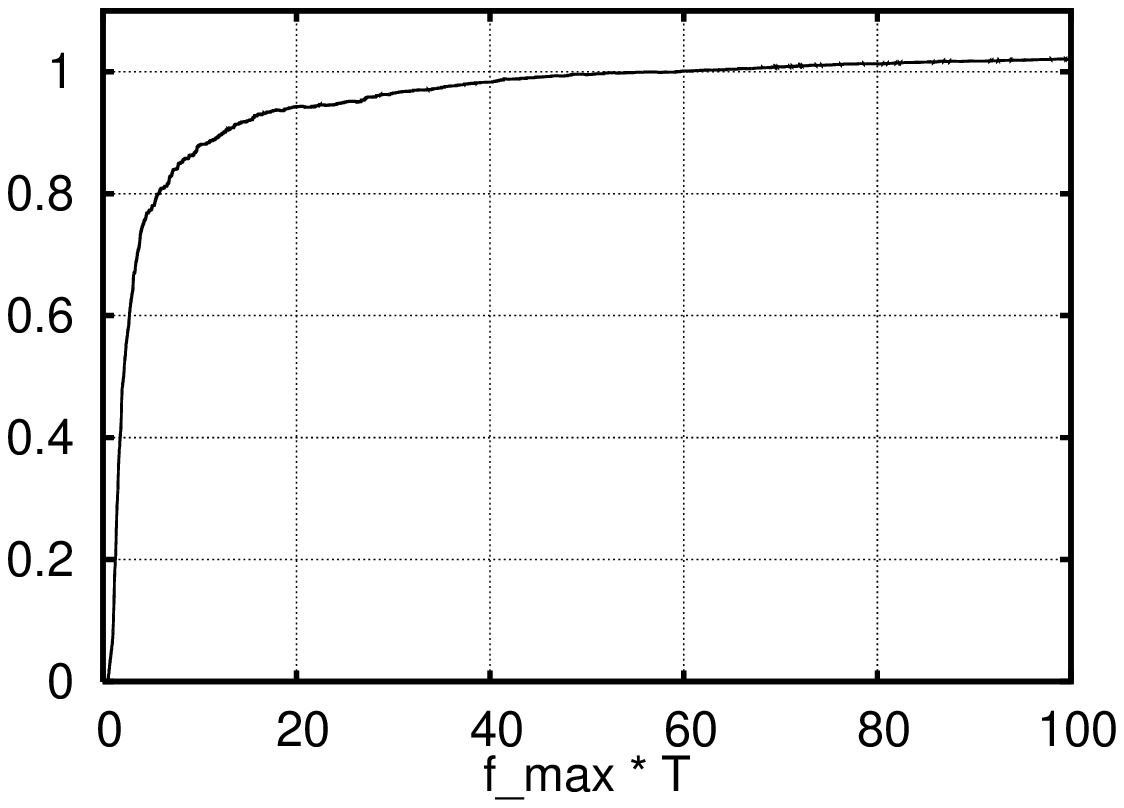} \\
\end{tabular}
\caption{The figure shows the precision of the alias-free approximation of the factor
$A(f_{\rm max})$. Top-left panel: the graph of the ratio of the derivative $A'(f_{\rm
max})$ (the inner integral in~(\ref{A-gen})) to its alias-free approximation
$2\pi^{n+0.5}\alpha_n T_{\rm eff}$. On an almost horizontal graph, we can see a sequence
of strong but narrow splashes corresponding to aliasing periods. Bottom-left panel: the
similar graph for the function $A(f_{\rm max})$ itself. The splashes at the aliasing
frequencies exist but are very small and do not produce significant perturbations. The
data were obtained for $n=3$ and $d_{\mathcal H}=1$ (with a free constant term in the
model $\mu_{\mathcal H}$). The $N=100$ timings of the mock input time series were
periodically gapped with a frequency corresponding to $Tf\approx 28.5$. At this gapping
frequency, the folded phases spanned only $\approx 10\%$ of the full period. Right panels
show similar graphs for the case of $N=30$ randomly spaced observations, $n=5$ and
$d_{\mathcal H}=1$.}
\label{fig_A}
\end{figure*}

Formally, the approximation~(\ref{A}) was based on certain assumptions of negligible
aliasing, which we have discussed above. Nevertheless, it was demonstrated in
\citep{Baluev08a} for the LS periodogram, that this approximation of the factor $A(f_{\rm
max})$ is quite precise in practice, even when the aliasing effects are strong. We may
expect the same behaviour of $A(f_{\rm max})$ for multi-harmonic periodograms. This is due
to the integral character of the representation~(\ref{A-gen}). Indeed, the aliasing may
result in a strong perturbation of the eigenvalues $\lambda_k$ and hence of the inner
integral in~(\ref{A-gen}). However, these perturbing effects are locked in very narrow
frequency intervals of the typical width $\Delta W \sim 1$. After integration over a wide
frequency range with $W\gg 1$, the resulting perturbation in the whole integral appear
insignificant. This is illustrated in Fig.~\ref{fig_A}.

Therefore, the only practically important source of a possible inaccuracy of the analytic
$\FAP$ estimation lies in the possible unsharpness of the Davies bound~(\ref{Dav}) itself
and in the possible inaccuracy of the associated alias-free
approximation~(\ref{alias-free}) itself.

\bsp

\label{lastpage}

\end{document}